\documentclass[letter,11pt]{article}
\pdfoutput=1

\usepackage{jheppub-mod}

\usepackage[T1]{fontenc}
\usepackage{amsthm}

\usepackage{listings}
\usepackage{mathrsfs}

\usepackage{xcolor}
\usepackage{graphicx, subcaption}

\hypersetup{pdftitle={Page Curves and Entanglement Islands for the Step-Function Vaidya Model of Evaporating Black Holes }, pdfauthor={Chang-Zhong Guo, Fu-Wen Shu}, citecolor=blue,linkcolor=blue,urlcolor=blue}

\newcommand{\be}{\begin{equation}}
\newcommand{\ee}{\end{equation}}

\usepackage{color}
\definecolor{purple}{rgb}{0.5,0,0.5}

\title{Page curves and Entanglement Islands for the Step-Function Vaidya Model of Evaporating Black Holes}

\author[a]{ Chang-Zhong Guo,}
\author[b]{ Wen-Cong Gan,}
\author[a,c,d*]{ Fu-Wen Shu}

\affiliation[a]{Department of Physics, Nanchang University, Nanchang, 330031, China}
\affiliation[b]{GCAP-CASPER, Physics Department, Baylor University, Waco, Texas 76798-7316, USA}
\affiliation[c]{Center for Relativistic Astrophysics and High Energy Physics, Nanchang University, Nanchang, 330031, China}
\affiliation[d]{Center for Gravitation and Cosmology, Yangzhou University, Yangzhou, China}

\emailAdd{chang.zhong.guo1997@gmail.com}
\emailAdd{Wen-cong$\_$Gan1@baylor.edu}
\emailAdd{shufuwen@ncu.edu.cn}

\abstract{It was proposed recently that the fine-grained entropy of the Hawking radiation can be expressed by the semiclassical island formula, which reproduces the unitary Page curve. In this paper, we choose the ``in'' vacuum state and apply the quantum extremal surface construction to study the Page curve for the step-function Vaidya model of evaporating black holes in four dimensions, which is produced by the spherical null shells. Metrics of the three regions of this spacetimes are obtained. In addition, the entanglement islands for the step-function Vaidya model of evaporating black holes at very late times are studied. When cutoff surface $A$ is located in Minkowski region III with $u_A < u_H$ at very late times, we find that the location of the boundary of island $\partial I$ depends on the value of $8M-v_A+v_I$.  Specifically, $\partial I$ is inside, at or outside the horizon when $8M-v_A+v_I$ is less than, equal to or larger than zero respectively. Moreover, when cutoff surface $A$ is located in Minkowski region III with $u_A > u_H$ after the black hole evaporates completely, we find that entanglement island still exists and $\partial I$ is located on an equal-time Cauchy surface of the observer $A$ when $r_{(A)}^2\geq64G_N\kappa c $. }

\begin{document}
\maketitle
\flushbottom

\section{Introduction}

 In 1970s, Hawking discovered that radiation of an evaporating black hole is the usual thermal radiation without carrying any information of the black hole \cite{Hawking:1974rv,Hawking:1975vcx,Hawking:1976ra}. Since then, the black hole information loss paradox has been one of the most fundamental problems in general relativity and quantum field theory for over 40 years. Hawking's calculation implies that the entanglement entropy of Hawking radiation will increase monotonically with time, it seemingly indicates that this process has violated the unitarity principle of quantum mechanics. The unitary evolution demands that the entanglement entropy of the Hawking radiation is portrayed by the Page curve \cite{Page:1993wv}. If the Page curve of the Hawking radiation can be reproduced, then the
black hole information paradox can be resolved. Recently, great breakthrough in solving black hole information paradox has been made via the semiclassical method
called \textit{island rule} \cite{Almheiri:2019hni,Penington:2019npb,Almheiri:2019psf,Almheiri:2020cfm}.

The holographic calculation of the entanglement entropy in QFT is originally from the Ryu-Takayanagi (RT) formula \cite{Ryu:2006bv,Hubeny:2007xt}, subsequent works generalized the RT formula to the quantum extremal surface (QES) prescription with the quantum corrections of the bulk fields \cite{Engelhardt:2014gca}. It is shown that the entropy of Hawking radiation can be calculated by the island formula, which can reproduce the Page curve of Hawking radiation in semiclassical gravitational calculation. Island rule states that the fine grained entropy of Hawking radiation is given by
\begin{equation}\label{island}
S(R)=\text{min}\left\{\text{ext}\left[\frac{\text{Area}(\partial I)}{4G_N}+S_{\text{ matter }}(I\cup R)\right]\right\},
\end{equation}
where $R$ is the region for collecting Hawking radiation, $I$ is the island which penetrates into the interior of the black hole, and the quantum extremal surface $\partial I$ is the boundary of the island. The first term in eq.(\ref{island}) is the classical area term from Ryu-Takayanagi (RT) extremal surface formula, and the second term is the bulk entanglement entropy of union of the island and the region $R$. Island rule is to extremize the generalized entropy for any possible quantum extremal surface and then take the one that results in the minimal generalized entropy. The proposed island formula can be further derived from the Euclidean path integral by making use of replica trick \cite{Penington:2019kki,Almheiri:2019qdq}.

Island formula was initially used to calculate the Page curve of an evaporating black hole in $2D$ Jackiw-Teitelboim (JT) gravity spacetime coupled to a thermal bath \cite{Almheiri:2019psf,Chen:2019uhq,Hollowood:2020cou}, and then the semiclassical method was extended to the case of two dimensional dilaton gravity in asymptotically flat spacetimes \cite{Anegawa:2020ezn,Hartman:2020swn,Tian:2022pso}. In two dimensional gravity models, there exist analytical solutions to the backreaction of the radiation under semiclassical approximations, and the expression for the entanglement entropy of matter fields in $2D$ conformally spacetime is well known in CFT method. In $D\geq3$ cases, many interesting and meaningful studies of the islands have been performed over the past few years, such as \cite{Saha:2021ohr,Yu:2021rfg} for BTZ black holes, \cite{Hashimoto:2020cas,Matsuo:2020ypv,Arefeva:2021kfx,Gan:2022jay,Du:2022vvg} for Schwarzschild black holes, \cite{Wang:2021woy,Kim:2021gzd,Yadav:2022fmo} for Reissner-Nordstr\"om black holes, \cite{Ling:2020laa} for the eternal black hole with charges on a
doubly-holographic model in general dimensions and \cite{He:2021mst} for general asymptotically flat eternal black holes. In spite of this, general method to calculate the entanglement entropy of matter fields in curved spacetimes larger than $2D$ is still missing. One possible strategy is to adopt certain approximations. For instance, it was argued in \cite{Hashimoto:2020cas,Matsuo:2020ypv} that one can calculate the entanglement entropy of matter fields in curved spacetimes by making use of $s$-wave approximation, for $4D$ Schwarzschild eternal black hole. However, as found by our previous work \cite{Gan:2022jay}, $s$-wave approximation for eternal black hole is questionable. Fortunately, it can be used to the one-sided Schwarzschild black hole which is formed from collapsing null shell.  As suggested in \cite{Gan:2022jay}, the associated vacuum state of the Hawking radiation in this case should be ``in'' vacuum state.  However, our previous work \cite{Gan:2022jay} did not take the backreaction of the Hawking radiation into consideration. Therefore, the model considered in \cite{Gan:2022jay}, strictly speaking, is not an evaporating black hole.

In this paper, we would like to fill this gap partially. We will apply the island rule to study the entanglement entropy of Hawking radiation and Page curves for the step-function Vaidya model of evaporating black holes, which is formed from the dynamical gravitational collapsing of spherical null shells \cite{Hiscock:1980ze}. We choose the ``in'' vacuum state to describe the black hole produced by the dynamical gravitational collapse, because it contains no incoming thermal radiation coming in from the past null infinity $\mathcal{J}^-$ \cite{Fabbri:2005mw}. Furthermore, when the cutoff surface $A$ is far from horizon, $s$-wave approximation is valid for the ``in'' vacuum state of the step-function Vaidya model.

This paper is organized as follows. In section \ref{111}, we briefly introduce the step-function Vaidya model of an evaporating black hole. We obtain the spacetime metrics of the three regions in $(u,v)$ coordinates, then the expression for the generalized entropy of Hawking radiation can be written in terms of $(u,v)$ coordinates for convenience. In section \ref{222}, we present an approximate method to calculate the entanglement entropy of Hawking radiation in curved $4D$ spacetime by $s$-wave approximation. In section \ref{333}, we use island rule to calculate the entanglement entropy of Hawking radiation and find the location of $\partial I$. First we will review our previous work \cite{Gan:2022jay} in subsection \ref{444}, we find that island $I$ emerges at late times and saves the entropy bound. When the observer is located in region II (see the Fig.\ref{vaidya1}) and far from the horizon, $\partial I$ is near and inside the horizon. In subsection \ref{555}, we discuss the case in which cutoff surface $A$ is located in the Minkowski region III with $u_A < u_H$ at very late times and find that the location of $\partial I$ depends on the value of $8M-v_A+v_I$. In subsection \ref{666}, we discuss the case in which cutoff surface $A$ is located in the Minkowski region III with $u_A > u_H$ after the black hole evaporates completely. We find that there exists an island in the Minkowski region III and $\partial I$ is located on an equal time Cauchy surface with the observer $A$ when $r_{(A)}^2\geq64G_N\kappa c $. The conclusion and discussion are in section \ref{777}.

\section{Step-function Vaidya model of an evaporating black hole}\label{111}

\subsection{The ingoing Vaidya metric for the model of collapse and evaporation}

\begin{figure}
\centering
\includegraphics[scale=0.550]{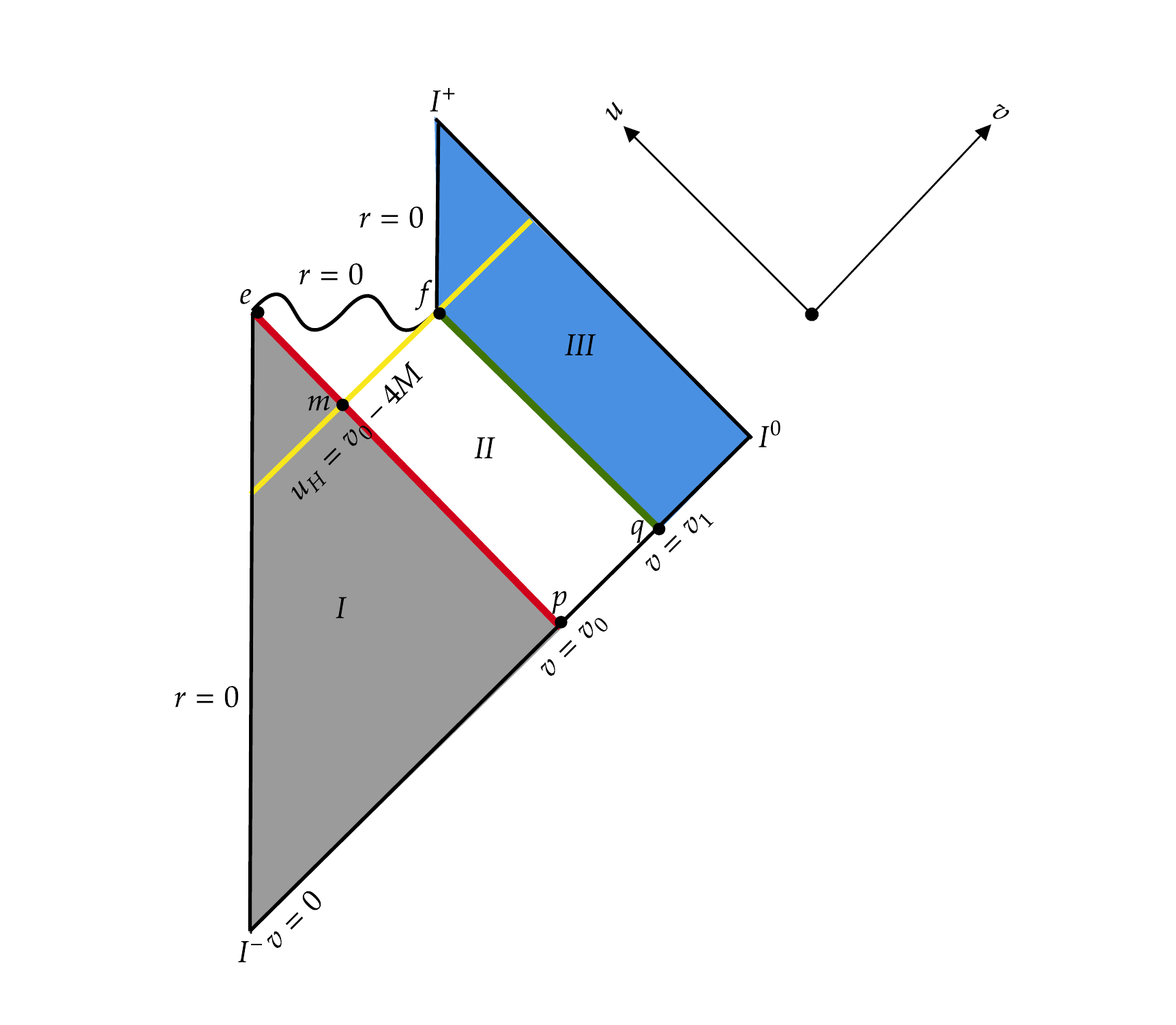}
\caption{The Penrose diagram for the step-function Vaidya model of an evaporating black hole spacetime formed by a positive mass spherical null shell collapsing at $v=v_0$ and an equal negative mass spherical null shell collapsing at $v=v_1$. The gray region I is Minkowski space inside the spherical null shell, the white region II is Schwarzschild spacetime and the blue region III also is Minkowski space after the black hole evaporates completely. The red line is an ingoing $v=v_0$ null geodesic, and the green line is an ingoing $v=v_1$ null geodesic. The yellow outgoing null geodesic is the event horizon of the black hole, where $u_H=v_0-4M$ is the location of the horizon.} \label{vaidya1}
\end{figure}

In this paper, we consider a simple solution to the semiclassical Einstein's equations describing the formation and evaporation of a black hole. This is given by the step-function Vaidya metric \cite{Hiscock:1980ze,Vaidya:1951zz} with the line element
\begin{equation}
ds^2=-\left(1-\frac{2M(v)}{r}\right)dv^2+2dvdr+r^2d\Omega^2,
\end{equation}
with\footnote{$H(x)$ is the step function with the following properties: \begin{equation}
\frac{dH(x)}{dx}=\delta(x), \quad \text{ and } H(x)=\begin{cases}0, & x \leq 0^-  \\ 1, & x \geq 0^+\end{cases}.\nonumber
\end{equation} }
\begin{equation}
M(v)=M[H(v-v_0)-H(v-v_1)]=\begin{cases}0, & v \leq v_0 \\ M, & v_0<v<v_1 \\ 0, & v \geq v_1\end{cases}
\end{equation}
Before proceed, let us first give more details about this toy model. Generally speaking, this model can partially model the evaporation of a black hole. On one hand, $\langle T_{\mu\nu}\rangle$ can be calculated explicitly for the entire spacetime and this model can be analytically analysed (see the Appendix.\ref{88838} for more details).  The Schwarzschild spacetime region II is formed by a positive mass null shell collapsing at $v=v_0$, and the spacetime region I is flat in the interior of the spherical null shell. Then there is an equal but negative mass null shell collapsing at $v=v_1$, which wipes the Schwarzschild black hole and the Minkowski spacetime arises in region III. $\langle T_{\mu\nu}\rangle$ is finite and regular everywhere except at the curvature singularity in region II. And $\langle T_{UU}\rangle$ in region III was calculated by Hiscock in Ref.\cite{Hiscock:1980ze}. On the other hand, however, the backreaction of the stress tensor is only modeled by the negative mass null shell. The negative mass null shell will cause the reduction of the mass of the black hole, leaving the Minkowski spacetime in region III. In some ways, the whole physical picture can describe approximately the process of an evaporating black hole. We can understand that the spacetime becomes Minkowskian after the black hole evaporates completely. In this sense we say that the step-function Vaidya toy model can partially model a black hole from formation to evaporation (see the Fig.\ref{vaidya1} for more details).

We can obtain a two dimensional spacetime by taking the $\theta=\text{constant}, \phi=\text{constant}$ slice of the four dimensional spherically symmetric spacetime. In terms of the double-null coordinates, the spacetime's metric for the three regions can be written as \footnote{ Noting that the radial coordinate $r$ is defined differently in the three regions, we define $r_{I}$ to represent the radial coordinate of region I, $r_{II}$ for the region II and $r_{III}$ for the region III.}:
\begin{eqnarray}\label{88386}
ds^2 &=& -dudv, \quad v \leq v_0,\nonumber\\
ds^2 &=& -\left(1-\frac{2M}{r_{II}}\right)du^{\ast}dv, \quad v_0<v<v_1,\nonumber\\
ds^2 &=& -dUdv, \quad  v \geq v_1,
\end{eqnarray}
where
\begin{equation}\label{5}
u=v-2r_{I}\ ,\ u^{\ast}=v-2{r_{II}}^{\ast} \ ,\ U=v-2r_{III} \ ,\text{ and } {r_{II}}^{\ast}=r_{II}+2M\ln\left|\frac{r_{II}}{2M}-1\right|.
\end{equation}

We consider massless scalar fields in the step-function Vaidya spacetime, and we choose the canonical set of null coordinates $(u_{in},v_{in})$ to define the usual vacuum state on the past null infinity $\mathcal{J}^-$. The scalar field modes are plane waves with the form $e^{-i\omega v}$ near the past null infinity $\mathcal{J}^-$ in the in-vacuum state, since the step-function Vaidya model spacetime is asymptotically flat. Then we can get the relation \cite{Hiscock:1980ze}:
\begin{equation}\label{1}
v_{in}=v.
\end{equation}
Using the reflection boundary condition of these scalar field modes through $r_{I}=0$ in region I and combining the eq.(\ref{1}), then it gives
\begin{equation}\label{6}
u_{in}=u.
\end{equation}

\subsection{The spacetime metrics for the step-function Vaidya model in $(u,v)$ coordinates}

\subsubsection{The spacetime metric of the region II in $(u,v)$ coordinates}
We can get the differential relations of the coordinate that matches across $v=v_0$. The metric is smooth along the ingoing $v=v_0$ null geodesic, then we have the connection condition \cite{Fabbri:2005mw}:
\begin{equation}\label{29}
r_{I}(u,v_0) = r_{II}(u^{\ast},v_0).
\end{equation}
Using the continuity conditions of the coordinate between region I and region II and the connection condition (\ref{29}), then we can obtain the differential equation for the ingoing $v=v_0$ null geodesic:
\begin{eqnarray}\label{2}
\frac{du^{\ast}}{du} &=& \frac{\partial u^{\ast}}{\partial r_{II}^{\ast}}\cdot\frac{dr_{II}^{\ast}}{dr_{II}}\cdot\frac{\partial r_{II}}{\partial u}
=\frac{r_{II}}{r_{II}-2M}=\frac{\frac{v_0-u}{2}}{\frac{v_0-u}{2}-2M}=\frac{u-v_0}{u+4M-v_0},
\end{eqnarray}
where in the third term and the penultimate term we have used $u=v_0-2r_{II}$ along the ingoing $v=v_0$ null geodesic. Then the spacetime's metric of the region II can be written as:
\begin{eqnarray}\label{3}
ds^2=-\left(1-\frac{2M}{r_{II}}\right)\frac{du^{\ast}}{du}\cdot dudv=-\left(1-\frac{2M}{r_{II}}\right)\frac{u-v_0}{u+4M-v_0}dudv.
\end{eqnarray}
To obtain the expression for $r_{II}$ in $(u,v)$ coordinates, first we should solve the differential eq.(\ref{2}). We can transform the differential eq.(\ref{2}) into an integral along the ingoing $v=v_0$ null geodesic, and $p$ is the starting point of the integral (see the Fig.\ref{vaidya1} for more details). The integral along the $v=v_0$ null geodesic is:
\begin{eqnarray}\label{4}
\int^{u^{\ast}}_{u^{\ast}_p}du^{\ast}=\int^{u}_{u_p}\frac{u-v_0}{u+4M-v_0}du.
\end{eqnarray}
The result of the integral eq.(\ref{4}) is:
\begin{eqnarray}\label{85}
u^{\ast}-u^{\ast}_p=u-u_p+4M\ln\left|\frac{u_p+4M-v_0}{u+4M-v_0}\right|=u+4M\ln\left|\frac{4M-v_0}{u+4M-v_0}\right|,
\end{eqnarray}
where in the last term we have used $u_p=0$. To obtain the specific expression for $u^{\ast}$ in $(u,v)$ coordinates further on, next we should figure out the expression for $u^{\ast}_p$ in terms of $v_0$ coordinate. Using the eq.(\ref{5}), then we have
\begin{eqnarray}\label{7}
u^{\ast}_p=v_0-2r^{\ast}_p=v_0-2\left(r_p+2M\ln\left(\frac{r_p}{2M}-1\right)\right)=-4M\ln\left(\frac{v_0}{4M}-1\right),
\end{eqnarray}
where in the last term we have used $u_p=v_0-2r_p=0$, then $r_p=\frac{v_0}{2}$. $m$ is the point at the event horizon of the black hole, the coordinate is given by
\begin{eqnarray}
r_m=2M\ ,\ \Longrightarrow u_m=u_H=v_0-2r_m=v_0-4M.
\end{eqnarray}
For simplicity, we should transform the eq.(\ref{85}) into an equation without absolute values. Actually, there are two cases for the eq.(\ref{85}), which are inside the horizon and outside the horizon respectively. First, we analyze the case inside the horizon. While inside the horizon $u > u_m$, we can get rid of the absolute values of the eq.(\ref{85}) and then we have the smooth connecting condition between region I and region II \cite{Gan:2022jay,Fabbri:2005mw}
\begin{eqnarray}\label{8}
u > u_H \ ,\ u^{\ast} = u_{in}-4M\ln\left(\frac{u_{in}-v_0+4M}{4M}\right),
\end{eqnarray}
where we have used eq.(\ref{6}). Finally, for the case outside the horizon $u < u_m$, we can also obtain the smooth connecting condition by combining eq.(\ref{5}) and eq.(\ref{7}):
\begin{eqnarray}\label{9}
u < u_H \ ,\ u^{\ast} = u_{in}-4M\ln\left(\frac{v_0-4M-u_{in}}{4M}\right).
\end{eqnarray}

Our ultimate goal is to get the expression for $r_{II}$ in $(u,v)$ coordinates, then we should solve the following equation by combining the eq.(\ref{5}), eq.(\ref{8}) and eq.(\ref{9}):
\begin{eqnarray}\label{10}
u^{\ast} = v-2\left(r_{II}+2M\ln\left|\frac{r_{II}}{2M}-1\right|\right).
\end{eqnarray}
We can get the expression for $r_{II}$ in $(u,v)$ coordinates by solving the eq.(\ref{10}), and we find that the expression for $r_{II}$ inside the horizon is the same as it outside the horizon. The expression for $r_{II}$ in $(u,v)$ coordinates is given by \footnote{In this paper, we use the $u$ coordinate to represent the coordinate of the ``in'' vacuum state uniformly.}
\begin{eqnarray}\label{11}
r_{II} = 2M\left(1+W\left[\frac{v_0-4M-u}{4M} \cdot e^{-1+\frac{v-u}{4M}}\right]\right),
\end{eqnarray}
where $W[x]$ is Lambert W function, which is also named product logarithm \footnote{The definition domain of Lambert W function $W[x]$ is: $x\in(-\frac{1}{e},+\infty)$. At the point $e$, $u_e=v_e=v_0$, we can check that the value of Lambert W function in eq.(\ref{11}) is $W[-\frac{1}{e}]=-1$, then the eq.(\ref{11}) $r_e=0$ is just true. And the $(u,v)$ coordinates in region II belong to the definition domain of Lambert W function.}.

\subsubsection{The spacetime metric of the region III in $(u,v)$ coordinates}
We can also get the differential relations of the coordinate that matches across $v=v_1$. The metric is smooth along the ingoing $v=v_1$ null geodesic, then we have the connection condition \cite{Fabbri:2005mw}:
\begin{equation}\label{17}
r_{II}(u,v_1) = r_{III}(U,v_1).
\end{equation}
Using the continuity conditions of the coordinate between region II and region III and the connection condition (\ref{17}), then we can get the differential equation for the ingoing $v=v_1$ null geodesic:
\begin{eqnarray}\label{12}
\frac{du}{dU}&=&\frac{du}{du^{\ast}}\cdot\frac{du^{\ast}}{dU}=\frac{u+4M-v_0}{u-v_0}\cdot\frac{du^{\ast}}{dU}=
\frac{u+4M-v_0}{u-v_0}\cdot\frac{d(v_1-2r_{II}^{\ast})}{d(v_1-2r_{III})}\nonumber\\
&=&\frac{u+4M-v_0}{u-v_0}\cdot\frac{dr_{II}^{\ast}}{dr_{III}}=\frac{u+4M-v_0}{u-v_0}\cdot\frac{r_{II}}{r_{II}-2M}
=\frac{u+4M-v_0}{u-v_0}\cdot\frac{\frac{v_1-U}{2}}{\frac{v_1-U}{2}-2M}\nonumber\\
&=&\frac{u+4M-v_0}{u-v_0}\cdot\frac{v_1-U}{v_1-U-4M},
\end{eqnarray}
where we have used the eq.(\ref{2}) and  $U=v_1-2r_{II}$ along the ingoing $v=v_1$ null geodesic. Then the spacetime's metric of the region III can be written as:
\begin{eqnarray}\label{13}
ds^2=-\frac{dU}{du}\cdot dudv=-\frac{u-v_0}{u+4M-v_0}\cdot\frac{v_1-U-4M}{v_1-U}dudv.
\end{eqnarray}

To obtain the expression for $U$ in $(u,v)$ coordinates, first we should solve the differential eq.(\ref{12}). We can transform the differential eq.(\ref{12}) into an integral along the ingoing $v=v_1$ null geodesic, and $q$ is the starting point of the integral (see the Fig.\ref{vaidya1} for more details). The integral along the $v=v_1$ null geodesic is:
\begin{eqnarray}\label{14}
\int^{u}_{u_q}\frac{u-v_0}{u+4M-v_0}du=\int^{U}_{U_q}\frac{v_1-U}{v_1-U-4M}dU.
\end{eqnarray}
The result of the integral eq.(\ref{14}) is:
\begin{eqnarray}\label{15}
u-u_q+4M\ln\left|\frac{u_q+4M-v_0}{u+4M-v_0}\right|=U-U_q+4M\ln\left|\frac{U_q+4M-v_1}{U+4M-v_1}\right|.
\end{eqnarray}
To obtain the specific expression for $U$ in $(u,v)$ coordinates further on, next we should figure out the expression for $u_q$ and $U_q$ in terms of $v_0$ and $v_1$ coordinates. Using the continuity conditions of the $u$ coordinate along the past null infinity $\mathcal{J}^-$ in the Penrose diagram, then we have $u_p=u_q=0$. Put $u_q=0$ and $v_q=v_1$ into the eq.(\ref{11}), then we have the expression for $r_{II}(q)$ in region II:
\begin{eqnarray}\label{16}
r_{II}(q) = 2M\left(1+W\left[\frac{v_0-4M}{4M} \cdot e^{-1+\frac{v_1}{4M}}\right]\right).
\end{eqnarray}
Next we can obtain the expression for $U_q$ in terms of $v_0$ and $v_1$ coordinates by combining the eq.(\ref{5}), eq.(\ref{17}) and eq.(\ref{16}):
\begin{equation}\label{18}
U_q = v_1-4M\left(1+W\left[\frac{v_0-4M}{4M} \cdot e^{-1+\frac{v_1}{4M}}\right]\right).
\end{equation}
Our ultimate goal is to obtain the expression for $U$ in $(u,v)$ coordinates, then we should solve the eq.(\ref{15}) first. For simplicity, we should transform the eq.(\ref{15}) into an equation without absolute values. The points at the ingoing $v=v_1$ null geodesic are outside the horizon, then we have
\begin{equation}\label{19}
u\leq u_f=u_H=v_0-4M\  ,\text{ and } v_1-U=2r_{III}(U,v_1)=2r_{II}(u,v_1)\geq 2r_{II}(f)=4M.
\end{equation}
Using the eq.(\ref{19}) and $u_q=0$ , then we can transform the eq.(\ref{15}) into an equation without absolute values:
\begin{eqnarray}\label{20}
u+4M\ln\left(\frac{4M-v_0}{u+4M-v_0}\right)=U-U_q+4M\ln\left(\frac{U_q+4M-v_1}{U+4M-v_1}\right).
\end{eqnarray}
Finally, plug the expression for $U_q$ (\ref{18}) back to the eq.(\ref{20}), then we can solve equation and obtain the expression for $U$ in $(u,v)$ coordinates \footnote{We can check the eq.(\ref{21}) for $U$ in $(u,v)$ coordinates. At the point $q$, $u_q=0$, plug $u_q=0$ back into eq.(\ref{21}), then we have  $U_q = v_1-4M\left(1+W\left[\frac{v_0-4M}{4M} \cdot e^{-1+\frac{v_1}{4M}}\right]\right)$. We can find that it is in accord with the eq.(\ref{18}) for $U_q$.}:
\begin{eqnarray}\label{21}
U =v_1-4M-4M\cdot W\left[-\frac{(4 M+u-v_0)\cdot e^{-\frac{u}{4 M}+\frac{v_1}{4 M}-1}}{4 M}\right].
\end{eqnarray}

\section{Entanglement entropy for dynamical black holes: overview }\label{Rel homo}\label{222}

For the $2D$ quantum system at criticality, the continuum limit is a conformal field theory with central charge $c$. The renormalized entanglement entropy  \footnote{The entanglement entropy of a single interval in vacuum state has the short distance UV cutoff scale $\epsilon$, in this paper we omit the cutoff scale $\epsilon$ since the divergence can be absorbed by the renormalization of the Newton constant $G_N$ \cite{Almheiri:2019psf,Hashimoto:2020cas}.} of a single interval in vacuum state in flat spacetime can be calculated by Cardy formula \cite{Calabrese:2004eu,Calabrese:2009qy}:
\begin{equation}
S_{\text{ matter }}=\frac{c}{3}\log\ell,
\end{equation}
where $\ell$ is the length of the single interval on the line in vacuum state in flat spacetime.

For $D\geq3$ flat spacetime, the renormalized entanglement entropy of matter fields in vacuum state is given by \footnote{More generally, the $\kappa$ factor for matter fields is positive, and the factor $\kappa$ for a massless field in $4D$ spacetime is numerically evaluated by Casini \cite{Casini:2009sr}, $\kappa = 0.00554$ for the massless bosons and $\kappa = 0.00538$ for the massless fermions. }\cite{Hashimoto:2020cas,Casini:2009sr}
\begin{equation}\label{26}
S_{\text{ matter }}=-\kappa c \frac{\text{ Area }}{L^2},
\end{equation}
where $\kappa$ is a constant which is related to the matter fields and $L$ is the geodesic distance between the parallelly placed boundary surfaces of the two endpoints for the interval. Entanglement entropy is an observable physical quantity, and the ``Area'' term in eq.(\ref{26}) is the area of the cutoff surface where the observer is located at.

For general $D\geq3$ curved spacetime, the entanglement entropy of quantum fields is usually not known in the bulk space. And the entanglement entropy of conformal fields is given by Cardy formula only in the two dimensional conformally flat spacetime under Weyl transformation. After Weyl transformed from $ds^2=-dx^+dx^-$ to $ds^2=-e^{2\rho(x^+,x^-)}dx^+dx^-$, entanglement entropy in $2D$ conformally flat spacetime is transformed as \cite{Almheiri:2019psf,Guo:2022ivd}:
\begin{equation}\label{25}
S_{e^{2\rho}ds^2}=S_{ds^2}+\frac{c}{6}\sum_{endpoints}\ln(e^{\rho})\ , \text{ and } S_{ds^2}=\frac{c}{3}\log\ell
\end{equation}
However, the Hawking radiation in $D\geq3$ spherically symmetric spacetime can be described by two dimensional $s$-wave approximation for a distant observer. Then for the step-function Vaidya model of evaporating black holes, the dynamics of Hawking radiation can be described effectively by a two dimensional CFT and we can apply the entanglement entropy of $2D$ CFT to calculate approximately the entanglement entropy in curved $4D$ spacetime.

\subsection{Hawking radiation in $4D$ one-sided dynamical spherically symmetric spacetime and $s$-wave approximation}

We consider the simplest form of the matter fields to be quantized in $4D$ dynamical spherically symmetric spacetime. First we investigate the behavior of massless scalar fields in the Schwarzschild ``out'' region $v_0<v<v_1$, and the Schwarzschild metric is:
\begin{equation}
ds^2 = -f(r)dt^2 +f(r)^{-1}dr^2 +r^2d\Omega^2\  ,\text{ and } f(r)=1-\frac{2M}{r}.
\end{equation}
The action of the massless scalar field $\varphi$ is \cite{Parker:2009uva}
\begin{equation}\label{82}
S = \int\sqrt{-g} d^4x\left[-\frac{1}{2}g^{\mu\nu}\nabla_{\mu}\varphi\nabla_{\nu}\varphi\right].
\end{equation}
By varying the scalar field $\varphi$ in eq.(\ref{82}), we can get the Klein-Gordon (KG) equation of the field:
\begin{equation}\label{22}
\square \varphi =\nabla^{\mu}\nabla_{\mu}\varphi=g^{\mu\nu}\nabla_{\mu}\nabla_{\nu}\varphi=0.
\end{equation}
We can decompose the scalar field $\varphi$ into partial waves
\begin{equation}\label{23}
\varphi(t,r,\theta,\phi) = \sum_{\ell,m}\frac{\varphi_{\ell}(t,r)}{r}Y_{\ell m}(\theta,\phi),
\end{equation}
where $Y_{\ell m}$ is the spherical harmonic functions. Plug the eq.(\ref{23}) back to eq.(\ref{22}), then the equation of motion for the scalar field $\varphi$ is converted into a two dimensional equation for $\varphi_{\ell}(t,r)$:
\begin{equation}
\left(-\frac{\partial^2}{\partial t^2}+\frac{\partial^2}{\partial r^{\ast^2}}-V_{\ell}(r)\right)\varphi_{\ell}(t,r)=0,
\end{equation}
where $r^{\ast}$ is the tortoise coordinate defined as $dr^{\ast}=f(r)^{-1}dr$, and the effective potential $V_{\ell}(r)$ for each partial wave with angular momentum $\ell$ is given by
\begin{equation}\label{24}
V_{\ell}(r)=\left(1-\frac{2M}{r}\right)\left(\frac{\ell(\ell+1)}{r^2}+\frac{2M}{r^3}\right).
\end{equation}
From the eq.(\ref{24}), we can see that only the $s$-wave with $\ell=0$ can survive at $r\geq 2M$ since partial waves with $\ell >0$ are more likely back-scattered by the effective potential $V_{\ell}(r)$ \cite{Fabbri:2005mw,Kawai:2013mda,Harlow:2014yka}. Also we notice that the effective potential vanishes both at $r=+\infty$ and at the event horizon $r=2M$, then the fields in $4D$ dynamical spherically symmetric spacetime can be treated as effective $2D$ free massless scalar fields at $r=+\infty$ and $r=2M$. It is easy to see that the outgoing Hawking radiation from the horizon to a distant observer ($r\gg 2M$) is dominated by the $s$-wave modes, then entanglement entropy has the logarithmic term \cite{Headrick:2019eth} under the $s$-wave approximation and entanglement entropy of the Hawking radiation in $4D$ dynamical spherically symmetric spacetime can be described approximatively by the eq.(\ref{25}) of $2D$ CFT.

\subsection{Island formula in the step-function Vaidya model}

Island formula for the fine grained entropy of the Hawking radiation is given by the eq.(\ref{island}). To calculate the fine grained entropy of the Hawking radiation, first we should extremize the generalized entropy with respect to the coordinates of the island,
\begin{equation}\label{27}
S_{\text{gen}}=\frac{\text{Area}(\partial I)}{4G_N}+S_{\text{ matter }}(I\cup R).
\end{equation}

The ``in'' vacuum state that we consider is pure in this paper, then the quantum state on the Cauchy slice $I\cup B \cup R$ is pure, where the region $B$ is the interval whose two endpoints are the quantum extremal surface $\partial I$ and the cutoff surface $A$ respectively. In quantum information theory, we have
\begin{equation}\label{28}
S_{\text{matter}}(B)=S_{\text{matter}}(I\cup R).
\end{equation}
Plug the eq.(\ref{28}) back to the eq.(\ref{27}) and combine the eq.(\ref{island}), then we can obtain simple expressions for the generalized entropy and island formula for the fine grained entropy of the Hawking radiation respectively:
\begin{eqnarray}
S_{\text{gen}}&=&\frac{\text{Area}(\partial I)}{4G_N}+S_{\text{ matter }}(B)\label{33},\\
S(R)=\text{min}\left\{\text{ext}\left[S_{\text{gen}}\right]\right\}&=&\text{min}\left\{\text{ext}\left[\frac{\text{Area}(\partial I)}{4G_N}+S_{\text{ matter }}(B)\right]\right\}\label{97}.
\end{eqnarray}

\section{Island for the step-function Vaidya model of evaporating black holes}\label{333}

\subsection{The observer in region II and far from the horizon}\label{444}

In this section we will review the calculation of the Page curve for one-sided asymptotically flat black hole done in \cite{Gan:2022jay}, and we first consider the case that the observer is located in region II and far from the horizon. Next, we will discuss the case that the metric covers the whole region II. In order to calculate the Page curve for convenience, we choose the Kruskal coordinates that cover the interior and exterior of the black hole.

The Kruskal coordinates are defined as:
\begin{eqnarray}\label{89}
U_{\text{kru}}&=&-4M\cdot e^{-\frac{u^{\ast}}{4M}}, \; V=4M\cdot e^{\frac{v}{4M}}, \; \text{(Outside horizon)}\nonumber\\
U_{\text{kru}}&=&4M\cdot e^{-\frac{u^{\ast}}{4M}}, \; V=4M\cdot e^{\frac{v}{4M}}. \; \text{(Inside horizon)}
\end{eqnarray}
We take the $\theta=\text{constant}, \phi=\text{constant}$ slice of the four dimensional spherically symmetric spacetime, then the metric in region II is converted to
 \begin{eqnarray}
 ds^2=-e^{2\rho(U_{\text{kru}},V)}dU_{\text{kru}} dV=-\frac{2M e^{-r_{II}/(2M)}}{r_{II}}dU_{\text{kru}} dV.
 \end{eqnarray}
From the eq.(\ref{8}) and eq.(\ref{9}), we can get the expression of $u$ for $u^{\ast}$:
\begin{eqnarray}
 u&=&v_0-4M-4M\cdot W\left[e^{-1+\frac{v_0}{4M}-\frac{u^{\ast}}{4M}}\right], \; u < u_H ,\nonumber\\
 u&=&v_0-4M-4M\cdot W\left[-e^{-1+\frac{v_0}{4M}-\frac{u^{\ast}}{4M}}\right], \; u > u_H.
 \end{eqnarray}
 At late times $u^{\ast}\rightarrow+\infty$, we have\footnote{We have used the property of the Lambert W function: $W[x] \simeq x$ when $x \rightarrow 0$.}
 \begin{eqnarray}\label{30}
 u(u^{\ast})&\simeq & -4M\cdot e^{-\frac{u^{\ast}}{4M}}-4M+v_0, \; u < u_H ,\nonumber\\
 u(u^{\ast})&\simeq & 4M\cdot e^{-\frac{u^{\ast}}{4M}}-4M+v_0, \; u > u_H.
 \end{eqnarray}
Combining the eq.(\ref{89}) and eq.(\ref{30}), then at late times we have
\begin{eqnarray}\label{31}
u \simeq v_H+U_{\text{kru}},
\end{eqnarray}
where $v_H=v_0-4M$. From the eqs.(\ref{89}) and (\ref{31}), we can easily obtain the following differential equation:
\begin{eqnarray}\label{32}
\frac{dU_{\text{kru}}}{d u} \simeq 1, \; \frac{dV}{d v}=\frac{V}{4M}.
\end{eqnarray}
In region II, the conformal factor $e^{2\rho(u,v)}$ in $(u,v)$ coordinates in the eq.(\ref{3}) is related to the conformal factor $e^{2\rho(U_{\text{kru}},V)}$ in Kruskal coordinates by isometry transformation:
\begin{eqnarray}
 ds^2 &=& -e^{2\rho(u,v)}dudv = -e^{2\rho(U_{\text{kru}},V)}dU_{\text{kru}} dV,\\
 \Longrightarrow e^{2\rho(u,v)}&=&e^{2\rho(U_{\text{kru}},V)}\cdot\frac{dU_{\text{kru}}}{d u}\frac{dV}{d v}\approx\frac{2M\cdot e^{-r_{II}/2M}}{r_{II}}\frac{V}{4M}\label{35},
 \end{eqnarray}
 where we have used the eq.(\ref{32}).

\subsubsection{With island and QES inside the horizon}

\begin{figure}
\centering
\includegraphics[scale=0.550]{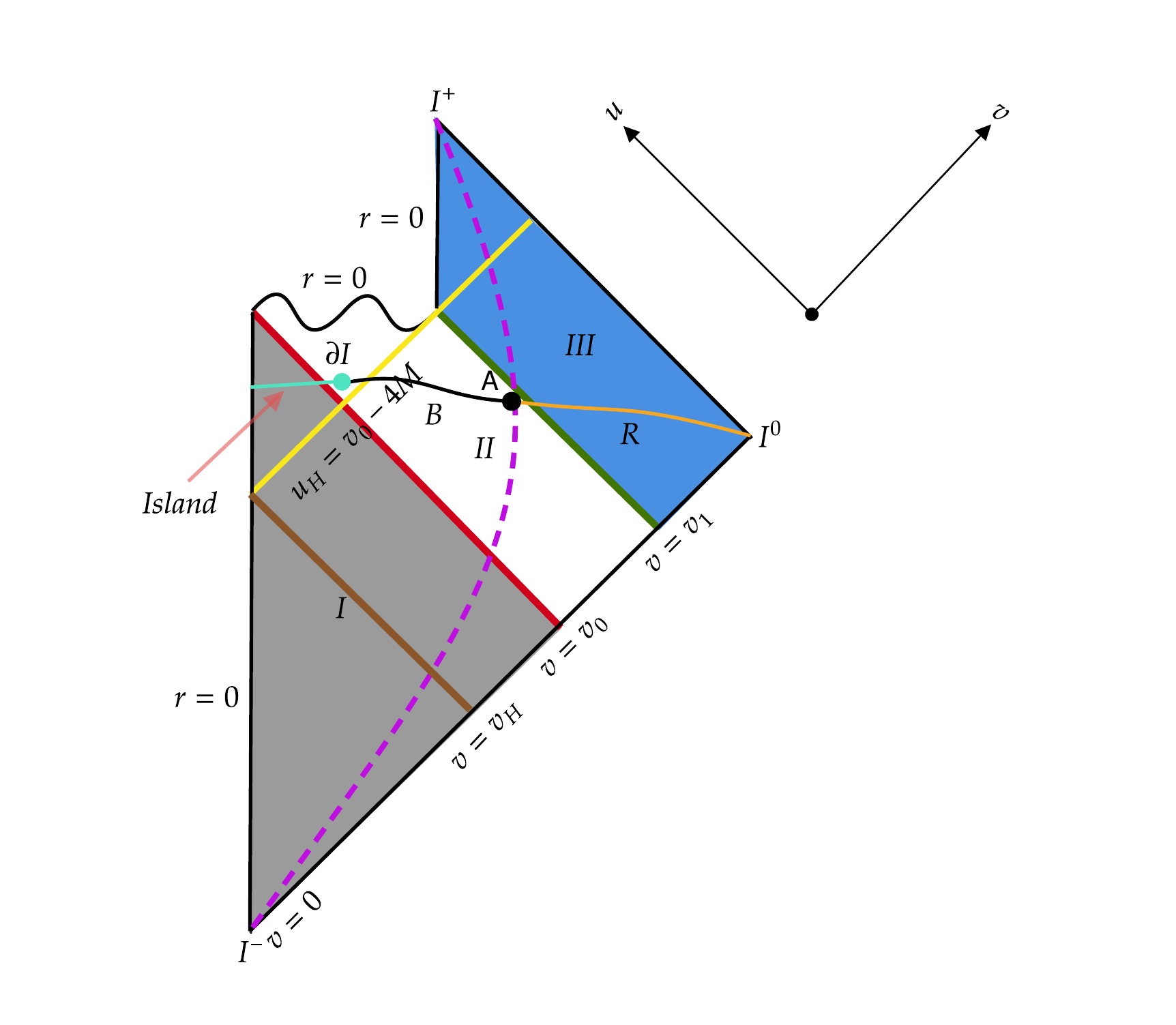}
\caption{The Penrose diagram for the step-function Vaidya model of an evaporating black hole spacetime with island and QES inside the horizon. The brown line is an ingoing $v=v_H=v_0-4M$ null geodesic. The observer who collects Hawking radiation in region $R$ (apricot yellow line) is located on cutoff surface $A$ (purple line) in region II. The cyan line is the island which penetrates into the interior of the black hole. The cyan dot $\partial I$ is the endpoint for the island, which is called quantum extremal surface (QES). The region $I\cup B\cup R$ is the whole Cauchy slice.} \label{observer2}
\end{figure}

In this section, we will use the semiclassical island formula to calculate the fine-grained entropy of Hawking radiation. We consider the case that the observer on the cutoff surface $A$ is far from the horizon at late times in region II, then the $s$-wave approximation is valid and entanglement entropy of Hawking radiation can be described approximatively by the eq.(\ref{25}) of $2D$ CFT. The construction is shown in Fig.\ref{observer2} for more details.

Combining the eq.(\ref{25}) for entanglement entropy of massless scalar fields in $2D$ conformally flat spacetime, the generalized entropy (\ref{33}) of Hawking radiation in this construction is given by \cite{Gan:2022jay}
\begin{equation}\label{34}
S_{\text{gen}}=\frac{\pi r_{(I)}^2}{G_N}+ \frac{c}{6}\ln\left(d(A,I)^2 e^{\rho_A} e^{\rho_I}\right),
\end{equation}
where $d(A,I)$ is the distance between $A$ and $\partial I$ in flat metric $ds^2=-dudv$ in vacuum state. The expression for $d(A,I)$ in terms of $(u,v)$ coordinates is given by
\begin{eqnarray}\label{36}
d(A,I)=\sqrt{[u_{A}-u_{I}][v_I-v_A]}.
\end{eqnarray}
Substituting eqs.(\ref{31}), (\ref{35}) and (\ref{36}) into the eq.(\ref{34}), then the expression for $S_{\text{gen}}$ can be given by
\begin{eqnarray}\label{37}
S_{\text{gen}}&=&\frac{\pi r_{(I)}^2}{G_N} +\frac{c}{12}\ln \left(
[v_H+U_{\text{kru(A)}}-(v_H+U_{\text{kru(I)}})]^2[v_I-v_A]^2 \frac{ e^{-(r_{(A)}+r_{(I)})/2M}\cdot V_AV_I}{4r_{(A)}r_{(I)}}
\right)\nonumber\\
&=&\frac{c}{12} \ln \left(\frac{V_A V_I \left(U_{\text{kru(A)}}-U_{\text{kru(I)}}\right)^2 e^{\chi_I} \left(\ln \left(\frac{V_I}{V_A}\right)\right){}^2}{e^2\left(1-\chi_I\right) \left(e^{W\left(-\chi_A\right)}-\chi_A\right)}\right)+\frac{4\pi M^2  \left(1-\chi_I\right)^2}{G_N},
\end{eqnarray}
where $\chi_i\equiv \frac{U_{\text{kru(i)}}V_i}{16eM^2},(i=A, I)$. In the second line of the eq.(\ref{37}), we have used the following equations \footnote{In the second term of the eq.(\ref{38}), we have assumed $r_{(I)}<2M$. For the case $r_{(I)}>2M$, we can use $r_{(I)} \approx 2M \left(1+e^{\frac{r_{(I)}^*}{2M}-1} \right)$ , then we will get the same expression of $S_{\text{gen}}$ as the eq.(\ref{37}).}:
\begin{equation}\label{38}
r_{(A)}=2M \left(1+W\left[e^{\frac{r_{(A)}^*}{2M}-1}\right] \right), \; r_{(A)} \approx 2M \left(1-e^{\frac{r_{(I)}^*}{2M}-1} \right),
\end{equation}
where $r_{(A)}^*$ and $r_{(I)}^*$ can be transformed into Kruskal coordinates via the eq.(\ref{29}).

To calculate the generalized entropy $S_{\text{gen}}$ easily and find where $\partial I$ is located at, first we assume $\partial I$ is near the horizon, then $r_{(I)}^{*} \rightarrow -\infty$ and $U_{\text{kru(I)}} \rightarrow 0$. Next we expand $S_{\text{gen}}$ to the first order of $U_{\text{kru(I)}}$, then we can obtain
\begin{eqnarray}\label{39}
&S_{\text{gen}}&=\left(\frac{4\pi  M^2}{G_N}+\frac{1}{12} c \ln \left(\frac{16 V_I U_{\text{kru(A)}} \chi_A M^2 \ln ^2\left(\frac{V_I}{V_A}\right)}{ e \left( e^{W\left(-\chi_A\right)}-\chi_A\right)}\right)\right)\nonumber\\
&&+U_{\text{kru(I)}} \left(\frac{1}{12} c \left(-\frac{2}{U_{\text{kru(A)}}}+\frac{V_I}{8 e M^2}\right)-\frac{\pi  V_I}{2 e G_N}\right)+O\left(U_{\text{kru(I)}}^2\right).
\end{eqnarray}
We can get the following equation by extremizing (\ref{39}) over $U_{\text{kru(I)}}$:
\begin{equation}
 \frac{\partial S_{\text{gen}}}{\partial U_{\text{kru(I)}}}\approx -\frac{c \left(16 e M^2-V_I U_{\text{kru(A)}}\right)}{96 e U_{\text{kru(A)}} M^2}-\frac{\pi  V_I}{2 e G_N}=0,
 \end{equation}
with the corresponding solution
\begin{equation}\label{41}
V_I= \frac{16 e c G_N M^2}{c G_N U_{\text{kru(A)}}-48 \pi  U_{\text{kru(A)}} M^2}.
\end{equation}
Also we can get the following equation by extremizing (\ref{39}) over $V_I$:
\begin{equation}\label{40}
 \frac{\partial S_{\text{gen}}}{\partial V_I}\approx \frac{1}{24} \left(\frac{2 c \left(1+\frac{2}{\ln \left(\frac{V_I}{V_A}\right)}\right)}{V_I}+\frac{U_{\text{kru(I)}} \left(\frac{c}{4M^2}-\frac{12 \pi }{G_N}\right)}{e}\right)=0.
\end{equation}
We can omit the term $\frac{2}{\ln \left(\frac{V_I}{V_A}\right)}$ in the eq.(\ref{40}) due to its smallness, then we have
\begin{equation}
\frac{U_{\text{kru(I)}} \left(\frac{c}{4M^2}-\frac{12 \pi }{G_N}\right)}{e}+\frac{2 c}{V_I}=0,
\end{equation}
with the corresponding solution
\begin{equation}\label{42}
 U_{\text{kru(I)}}=-\frac{8 e c G_N M^2}{V_I \left(c G_N -48 \pi  M^2\right)}.
\end{equation}

We can obtain the solutions from the eqs.(\ref{41}) and (\ref{42}):
\begin{equation}\label{43}
U_{\text{kru(I)}}=-\frac{U_{\text{kru(A)}}}{2}, \; V_I=\frac{16 e c G_N M^2}{c G_N U_{\text{kru(A)}}-48 \pi  U_{\text{kru(A)}} M^2}.
\end{equation}
The observer on the cutoff surface $A$ is outside the horizon with $U_{\text{kru(A)}}<0$, thus we have $U_{\text{kru(I)}}>0$ from the eq.(\ref{43}). The result means that $\partial I$ is inside the horizon. Moreover, we can also obtain the following equation from the eq.(\ref{43}) by using $c G_N \ll M^2$ :
\begin{equation}\label{44}
\frac{ U_{\text{kru(I)}} V_I}{16M^2}=\frac{e c G_N}{96 \pi M^2-2c G_N} \approx \frac{e c G_N}{96 \pi M^2}.
\end{equation}
In terms of Kruskal coordinates, we have $\frac{ U_{\text{kru(I)}} V_I}{16M^2}=e^{r_{(I)}^{*}/2M}$. When $\partial I$ is near the horizon with $r_{(I)}^{*} \rightarrow -\infty$, we can find that the result for $\frac{ U_{\text{kru(I)}} V_I}{16M^2}$ is coincided with (\ref{44}) and our assumption is true.

Finally, plug the solutions (\ref{43}) back to the generalized entropy in eq.(\ref{39}), then the generalized entropy  $S_{\text{gen}}$ can be reduced to
\begin{eqnarray}\label{45}
 S_{\text{gen}}&=&\frac{4\pi  M^2}{G_N}+\frac{1}{12} c \left(\ln \left(-\frac{12 c G_N \chi_A  M^2 \ln ^2\left(-\frac{e c G_N}{3 \pi  U_{\text{kru(A)}} V_A}\right)}{\pi  \left( e^{W\left(-\chi_A\right)}-\chi_A \right)}\right)-1\right)+O\left(G_N^1\right)\\
 &\approx& \frac{4\pi  M^2}{G_N}=S_{\text{BH}},
\end{eqnarray}
where we have used $c G_N \ll M^2$ in the penultimate term. At late time in region II, it is easily seen that the generalized entropy of Hawking radiation being a constant by considering the construction of the island. 

\subsubsection{Without island and the Page time}

\begin{figure}
\centering
\includegraphics[scale=0.550]{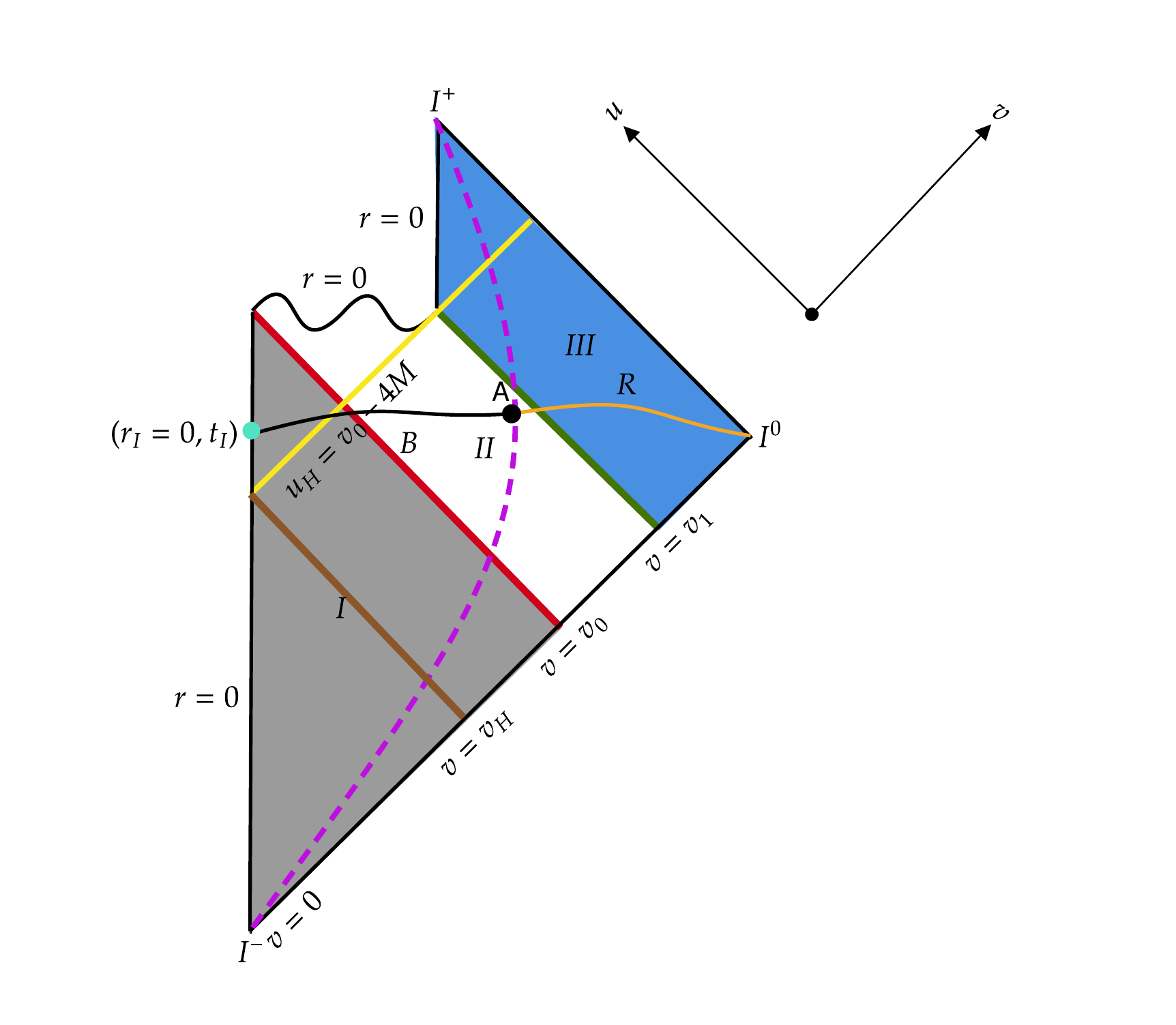}
\caption{The Penrose diagram for the step-function Vaidya model of an evaporating black hole spacetime without island. In this case, $r_I=0$ can be equivalently fixed.} \label{o1}
\end{figure}

In this section, we will calculate the radiation entropy at late times without island and the result will cause the black hole information problem. Besides, we assume that the $s$-wave approximation is valid in the Minkowski region I and then we can use the matter entropy formula of $2D$ CFT for calculating the entanglement entropy of Hawking radiation. In this case, we have $r_{(I)}=0$ and $t_{(I)}=\text{ constant }$ (see the Fig.\ref{o1} for more details). Then the gravity term $\frac{\text{Area}(\partial I)}{4G_N}$ in eq.(\ref{27}) is zero and the generalized entropy is only reduced to the matter field term (\ref{25}): $S_{\text{gen}}=S_{\text{matter}}(I\cup R)$. Now $r_{(I)}=0$ is in the Minkowski region I inside the spherical null shell, then we have $u(I)=t_{(I)}$ and $e^{2\rho(u,v)(I)}=1$. By using the eq.(\ref{28}), the matter sector $B$ of the radiation entropy contributes as
 \begin{eqnarray}\label{46}
S_{\text{gen}}&=&S_{\text{matter}}(B)=\frac{c}{12} \ln \left([v_H+U_{\text{kru(A)}}-t_{(I)}]^2[v_I-v_A]^2 \frac{2M e^{-r_{(A)}/2M}}{r_{(A)}}\frac{V_A}{4M}
\right)\nonumber\\
&=&\frac{1}{12} c \ln \Bigg(\frac{2M}{r_{(A)}} \sqrt{\frac{r_{(A)}}{2M}-1} e^{-\frac{r_{(A)}+t_{(A)}}{4M}} \Bigg(2 e^{\frac{r_{(A)}}{4M}} \sqrt{2M \left(r_{(A)}-2M\right)}\nonumber\\
&+&e^{\frac{t_{(A)}}{4M}} \left(4M+t_{(I)}-v_0\right)\Bigg)^2\Bigg)+\frac{1}{6} c \ln \left( \left(2M \ln \left(\frac{r_{(A)}}{2M}-1\right)+r_{(A)}+t_{(A)}-t_{(I)}\right)\right)\nonumber\\
&\approx &\frac{c}{48M} t_{(A)},
\end{eqnarray}
where we have taken into account the late time limit in the last term that $t_{(A)} \gg r_{(A)} \gg 2M$. Obviously, the radiation entropy increases linearly with time at late time and it will eventually become greater than $S_{\text{BH}}$ for sufficiently large $t_{(A)}$. It is consistent with Hawking's result which violates the unitarity at late time, and the black hole information paradox appears.

Next, we will calculate the Page time which is defined as the moment when the radiation entropy of the whole system reaches the maximum. Compare the eq.(\ref{45}) with the eq.(\ref{46}) and use the eq.(\ref{97}), we have
\begin{equation}
S(R)=\text{min}( S_{\text{gen}}) \approx \text{min}(S_{\text{matter}}(B),S_{\text{BH}}).
\end{equation}
Then we can calculate the Page time:
\begin{equation}
 S_{\text{matter}}(B)=\frac{c}{48M} t_{(A)}=S_{\text{BH}}, \; \Longrightarrow t_{\text{Page}}\approx\frac{48 M}{c}S_{\text{BH}}.
\end{equation}

\subsection{The observer in region III with $u_A < u_H$}\label{555}

\begin{figure}
\centering
\includegraphics[scale=0.550]{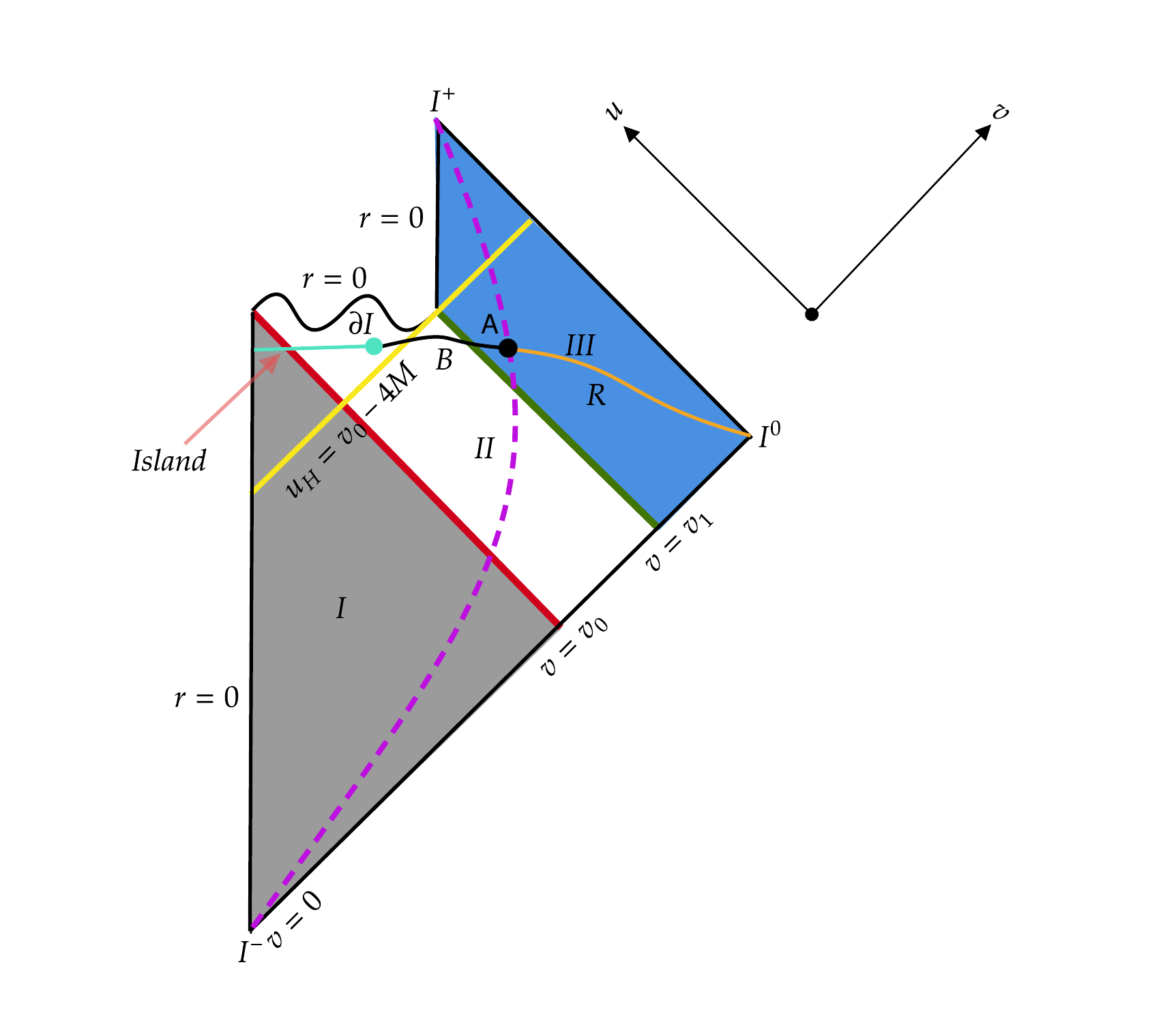}
\caption{The Penrose diagram for the step-function Vaidya model of an evaporating black hole spacetime with island. The observer is located in region III with $u_A < u_H$.} \label{observer3}
\end{figure}

In this section, we consider the case that the observer on the cutoff surface $A$ is located in the Minkowski region III with $u_A < u_H$ at very late times (see the Fig.\ref{observer3}). We assume that the $s$-wave approximation is valid in the Minkowski region III, then the generalized entropy of Hawking radiation can be given by the eq.(\ref{34}). First we make an assumption that $\partial I$ is near the horizon in the following calculations, then the expression for $r_{(I)}$ in $(u,v)$ coordinates can approximatively be written as \footnote{For the case $\partial I$ near the horizon, then $u_{I}\rightarrow v_0-4M$. From the eq.(\ref{11}), we have $r_{(I)} = 2M\left(1+W\left[\frac{v_0-4M-u_{I}}{4M} \cdot e^{-1+\frac{v_{I}-u_{I}}{4M}}\right]\right)$. Expand $r_{(I)}$ about the point $u_{I}=v_0-4M$ to the first order of $u_{I}+4M-v_0$, the we can obtain the eq.(\ref{47}).}
\begin{eqnarray}\label{47}
r_{(I)}\approx 2M\left(1+\frac{v_0-4M-u_{I}}{4M} \cdot e^{-1+\frac{v_{I}-u_{I}}{4M}}\right).
\end{eqnarray}
From the spacetime's metric of the region II in (\ref{3}), we can obtain the expression for the conformal factor $e^{2\rho_{I}}$ in $(u,v)$ coordinates:
\begin{eqnarray}\label{48}
e^{2\rho_{I}}=\left(1-\frac{2M}{r_{(I)}}\right)\frac{u_{I}-v_0}{u_{I}+4M-v_0}.
\end{eqnarray}
Plug the expression for $r_{(I)}$ (\ref{47}) back to the eq.(\ref{48}), then put eqs.(\ref{36}),(\ref{47}) and (\ref{48}) into the eq.(\ref{34}), finally we can get the expression for the generalized entropy of Hawking radiation in $(u,v)$ coordinates \footnote{For simplicity, it is not necessary to calculate the conformal factor $e^{2\rho_{A}}$ for the observer on the cutoff surface $A$ in $(u,v)$ coordinates. The conformal factor $e^{2\rho_{A}}$ is a constant since the position of the observer on the cutoff surface $A$ in the spacetime is fixed.}:
\begin{eqnarray}\label{49}
S_{\text{gen}}&=&\frac{4 \pi  \left(M+\frac{v_0-4M-u_{I}}{4} \cdot e^{-1+\frac{v_{I}-u_{I}}{4M}}\right)^2}{G_N}\nonumber\\
&+&\frac{c}{12}\log\left[\frac{e^{2\rho_{A}} (u_{A}-u_{I})^2 (v_A-v_I)^2\cdot e^{\frac{v_{I}}{4M}}(u_{I}-v_0)  }{e^{\frac{v_I}{4 M}} (4 M+u_{I}-v_0)-4M\cdot e^{\frac{u_{I}}{4 M}+1}}\right].
\end{eqnarray}

Next we expand $S_{\text{gen}}$ about the point $u_{I}=v_0-4M$ to the first order of $u_{I}+4M-v_0$, then we can obtain
\begin{eqnarray}\label{50}
S_{\text{gen}}&=&\left(\frac{4 \pi  M^2}{G_N}+\frac{c}{12}  \log \left(e^{2\rho_{A}}e^{\frac{v_I-v_0}{4 M}} (v_A-v_I)^2 (4 M+u_A-v_0)^2 \right)\right)\nonumber\\
&+&(4 M+u_I-v_0)\cdot \Bigg(-\frac{c e^{-\frac{v_0}{4 M}} \Bigg((2 u_A-2 v_0+16M) e^{\frac{v_0}{4 M}}+(v_0-4M-u_A) e^{\frac{v_I}{4 M}}\Bigg)}{48 M \cdot(4 M+u_A-v_0)}\nonumber\\
&-&\frac{2 \pi  M\cdot e^{\frac{v_I-v_0}{4 M}}}{G_N}\Bigg)+O\left((4 M+u_I-v_0)^2\right).
\end{eqnarray}
Extremizing $S_{\text{gen}}$ (\ref{50}) over $u_I$, we have
\begin{eqnarray}\label{51}
\frac{\partial S_{\text{gen}}}{\partial u_I}\approx \frac{ \left(-\frac{c \left(e^{\frac{v_I}{4 M}} (-4 M-u_A+v_0)+2 e^{\frac{v_0}{4 M}} (8 M+u_A-v_0)\right)}{4 M+u_A-v_0}-\frac{96 \pi  M^2\cdot e^{\frac{v_I}{4 M}}}{G_N}\right)}{48 M\cdot e^{\frac{v_0}{4 M}}}=0.
\end{eqnarray}
The solution of $v_I$ to the eq.(\ref{50}) is
\begin{equation}\label{52}
 v_I=4M\left(2i\pi c_1+\log \left[\frac{2 c G_N e^{\frac{v_0}{4 M}} (8 M+u_A-v_0)}{\left(c G_N-96 \pi  M^2\right) (4 M+u_A-v_0)}\right]\right), \; c_1\in \mathbb{Z},
 \end{equation}
 where $\mathbb{Z}$ represents integers. In fact, we should drop the complex solution since the $(u,v)$ coordinates can only be real numbers, then we are left with
\begin{equation}\label{53}
 v_I=4M\cdot\log \left[\frac{2 c G_N e^{\frac{v_0}{4 M}} (8 M+u_A-v_0)}{\left(c G_N-96 \pi  M^2\right) (4 M+u_A-v_0)}\right].
\end{equation}
Extremizing $S_{\text{gen}}$ (\ref{50}) over $v_I$, then we can get
\begin{eqnarray}\label{54}
\frac{\partial S_{\text{gen}}}{\partial v_I}& \approx & \frac{e^{-\frac{v_0}{4 M}}}{192 G_N M^2 (v_I-v_A)}\cdot  \Bigg(c G_N  \Bigg(e^{\frac{v_I}{4 M}} (v_I-v_A) (4 M+u_I-v_0)\nonumber\\
&+&4 M e^{\frac{v_0}{4 M}} (8 M-v_A+v_I) \Bigg)- 96 \pi  M^2 e^{\frac{v_I}{4 M}} (v_I-v_A) (4 M+u_I-v_0) \Bigg)=0,
\end{eqnarray}
with the corresponding solution
\begin{eqnarray}\label{55}
u_I&=&v_0-4M +\frac{4cG_NM\cdot e^{\frac{v_0-v_I}{4 M}}(8M-v_A+v_I) }{(v_I-v_A)\cdot(96\pi M^2-cG_N)}\nonumber\\
&=&u_H +\frac{4cG_NM\cdot e^{\frac{v_0-v_I}{4 M}}(8M-v_A+v_I) }{(v_I-v_A)\cdot(96\pi M^2-cG_N)}.
\end{eqnarray}
Since $v_I < v_A$ and $96\pi M^2 \gg cG_N$, it is easy to see that the location of $\partial I$ depends on the value of $8M-v_A+v_I$ from the eq.(\ref{55}):
\begin{eqnarray}\label{56}
8M-v_A+v_I \begin{cases} <0, & u_I > u_H , \; \text{($\partial I$ is inside the horizon)}\\ =0, & u_I = u_H, \; \text{($\partial I$ is at the horizon)} \\ >0, &u_I > u_H, \; \text{($\partial I$ is outside the horizon)}\end{cases}
\end{eqnarray}

We know that $v_I >0$ and $u_A < u_H=v_0-4M$ in the Penrose diagram, then we can obtain the following inequalities from the (\ref{53}):
\begin{equation}\label{57}
8 M+u_A-v_0>0, \;  \;\frac{2 c G_N e^{\frac{v_0}{4 M}} (8 M+u_A-v_0)}{\left(c G_N-96 \pi  M^2\right) (4 M+u_A-v_0)}>1,
\end{equation}
where we have used the condition for the domain of the logarithmic function in eq.(\ref{53}). From the Penrose diagram (\ref{observer3}), we can also get the following relational inequalities:
\begin{equation}\label{58}
v_A>v_1>v_I>v_0>0, \;  \;u_A>0.
\end{equation}
To see the relation between the changes of the $(u_A,v_0,v_A,M,cG_N)$ parameters and the location of $\partial I$ further on, next we will reduce the eq.(\ref{56}) to its simplest form by making indexes being dimensionless. We define these dimensionless parameters as follows:
\begin{equation}\label{59}
\alpha\equiv \frac{96\pi M^2}{cG_N}, \; \beta\equiv\frac{u_A}{4M}, \; \gamma\equiv\frac{v_0}{4M}, \;\eta\equiv\frac{v_A}{4M}.
\end{equation}

\begin{figure}[htbp]
        \centering
        \subcaptionbox{}{
        \includegraphics[width = .31\linewidth]{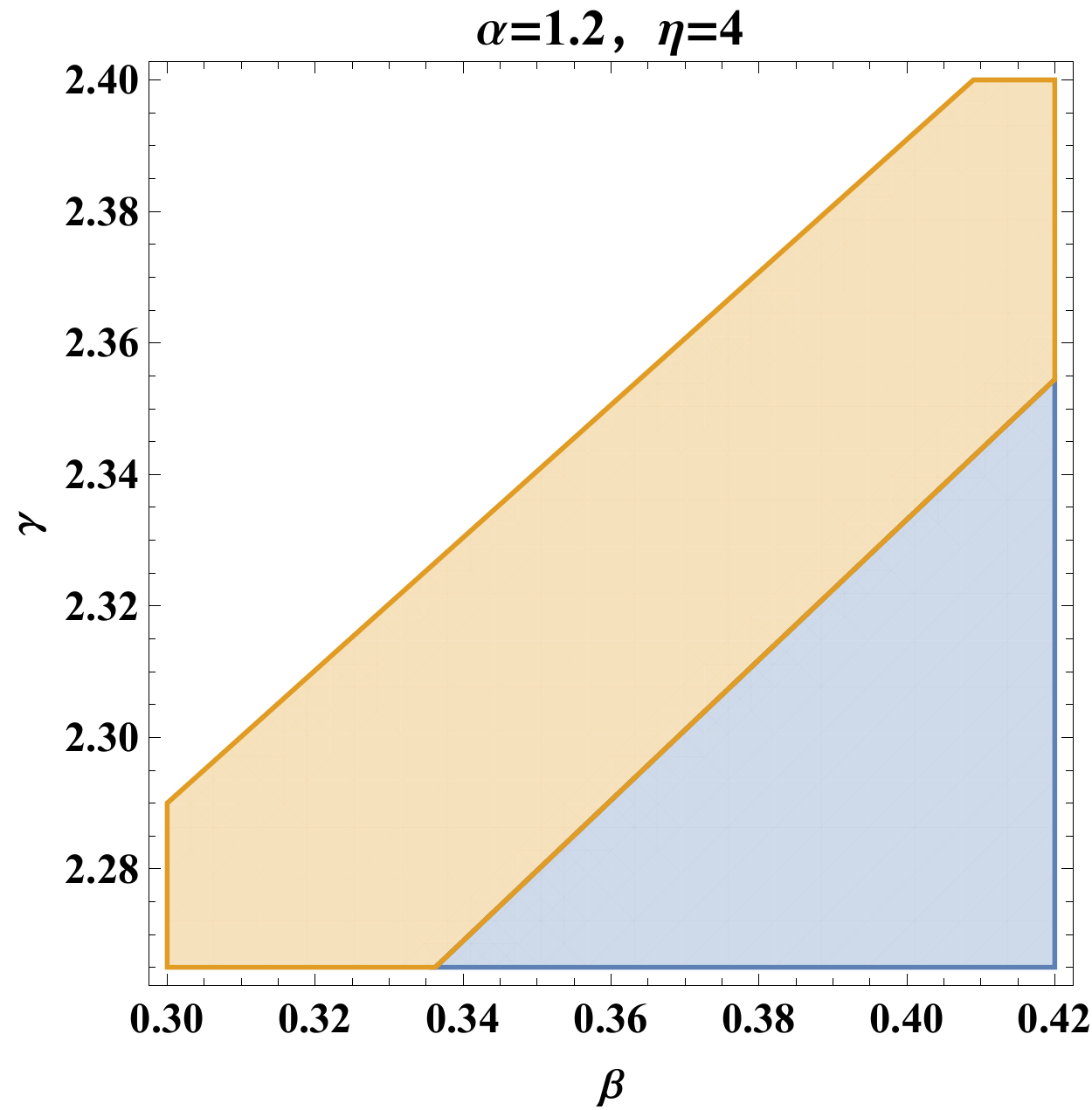}
    }
        \subcaptionbox{}{
        \centering
        \includegraphics[width = .31\linewidth]{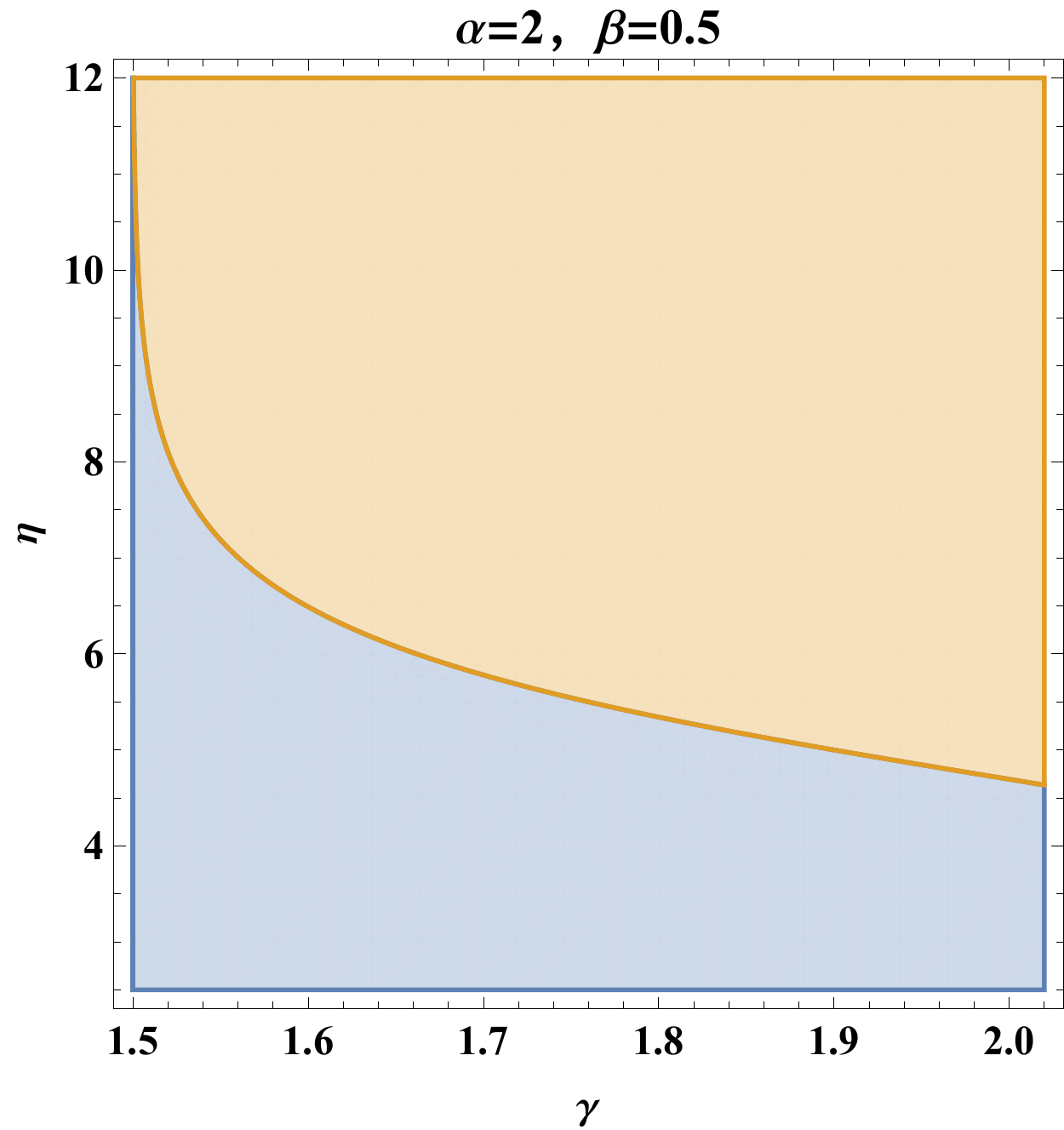}
    }
        \subcaptionbox{}{
        \includegraphics[width = .31\linewidth]{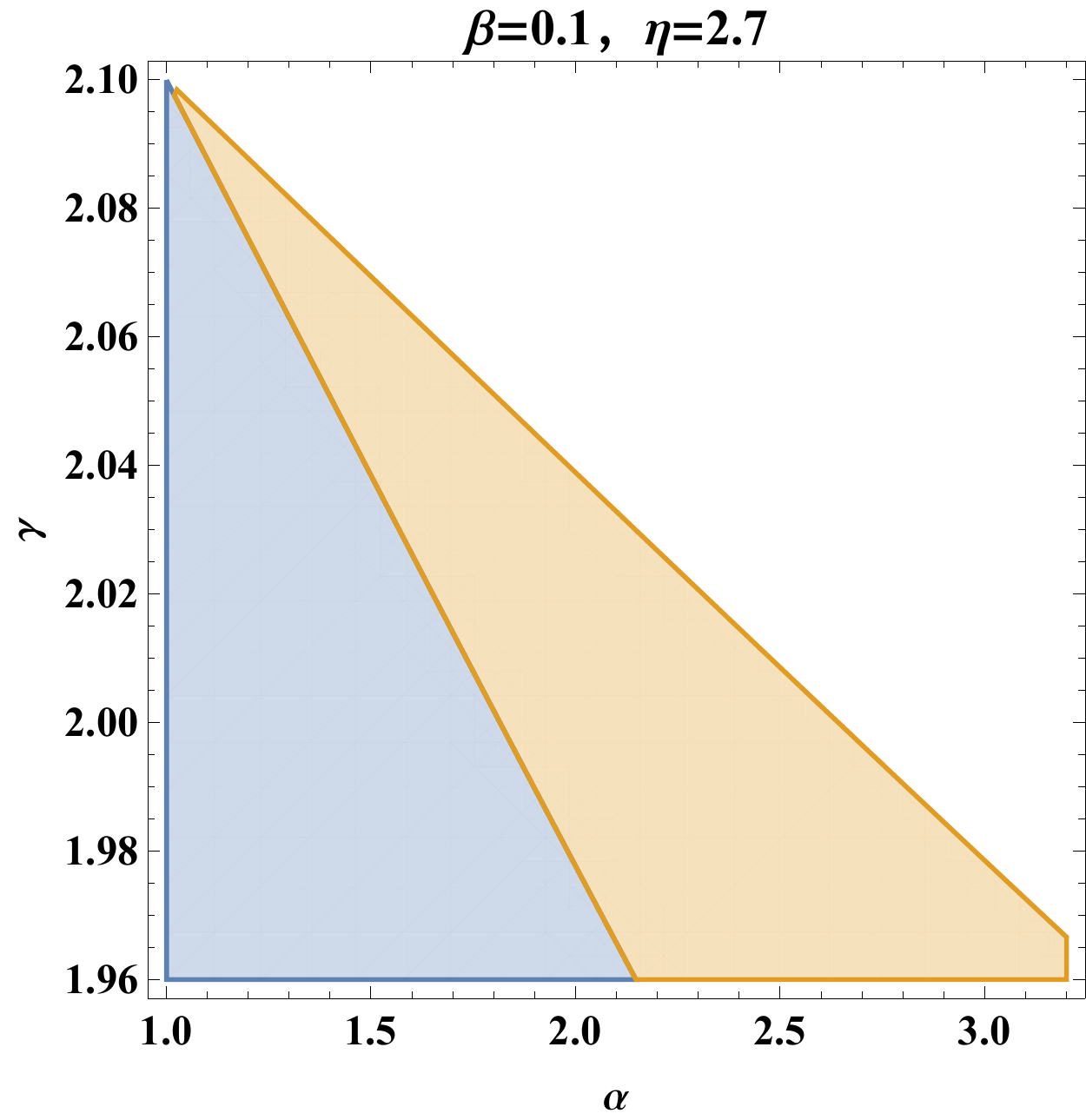}
    }
    \centering
        \subcaptionbox{}{
        \includegraphics[width = .31\linewidth]{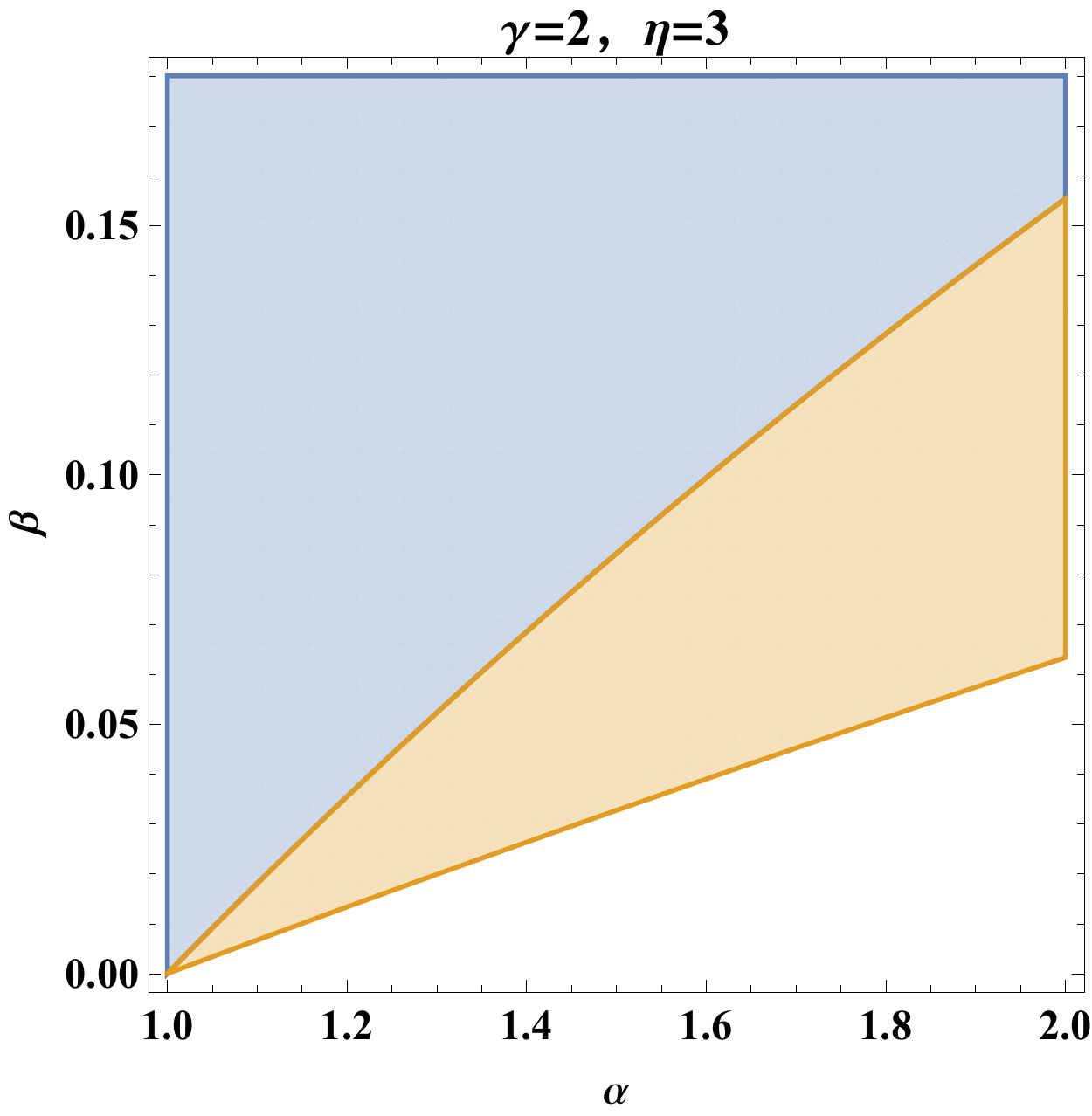}
    }
        \subcaptionbox{}{
        \centering
        \includegraphics[width = .31\linewidth]{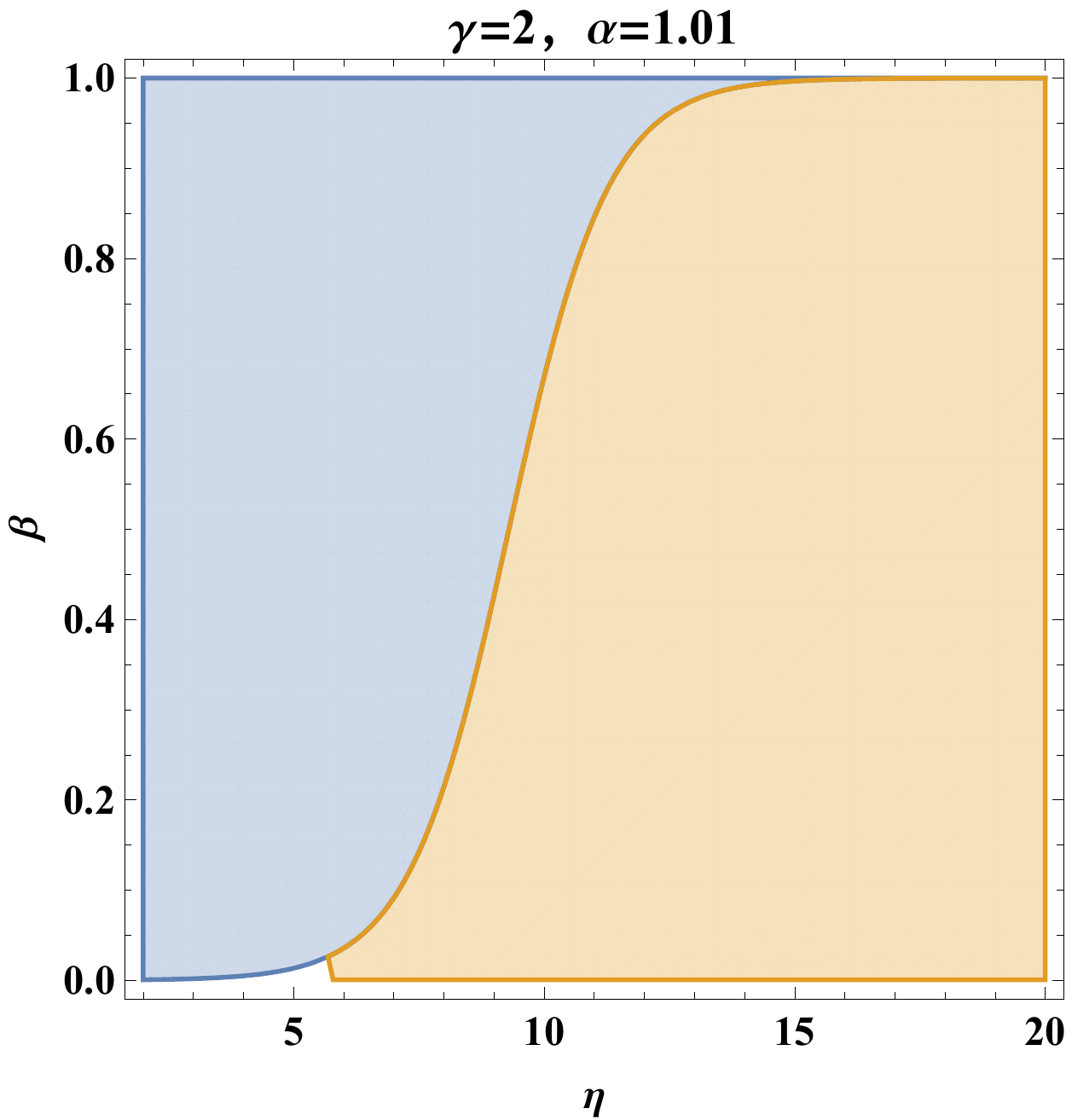}
    }
        \subcaptionbox{}{
        \includegraphics[width = .31\linewidth]{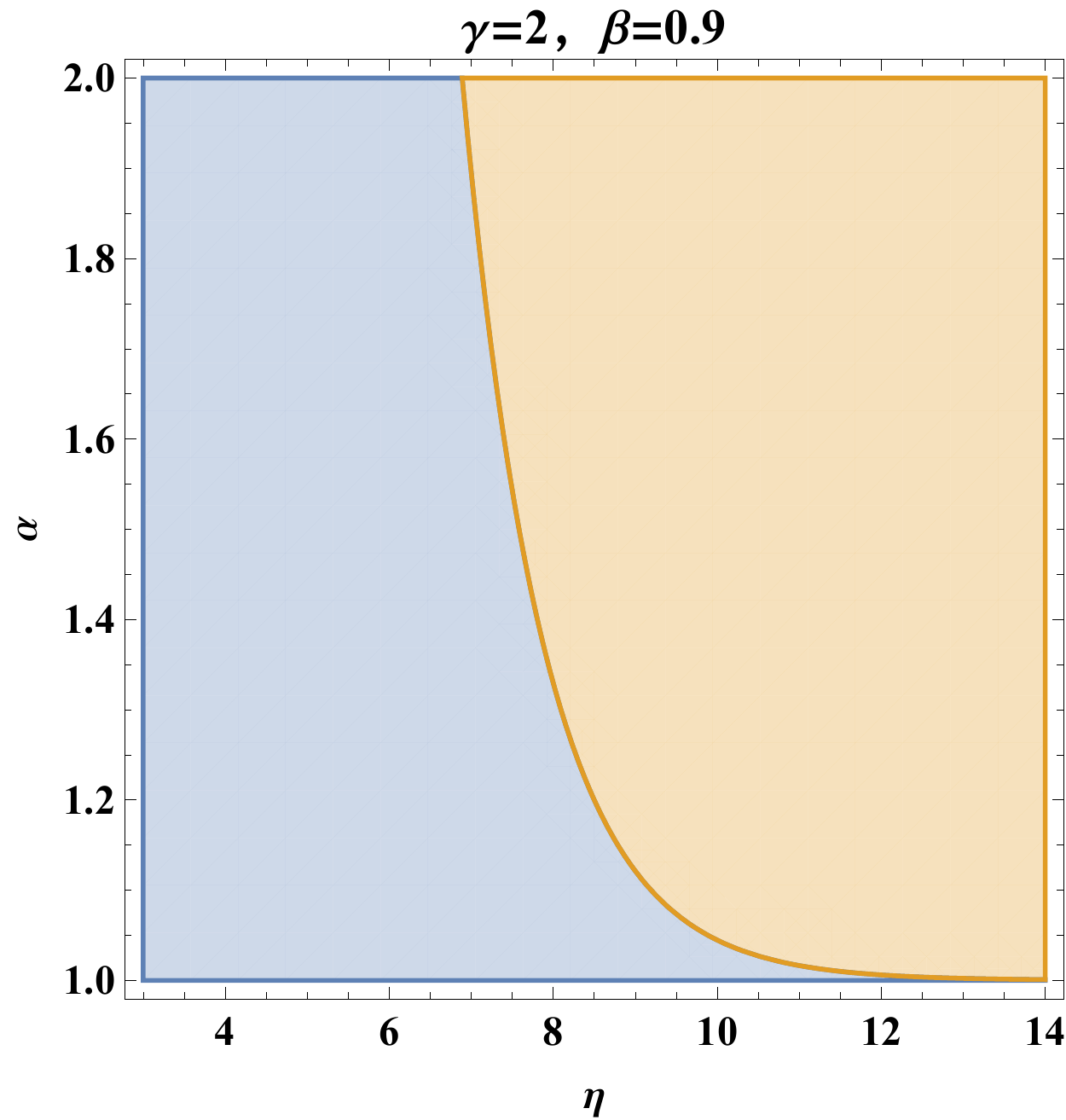}
    }
        \caption{ We fix two dimensionless parameters: $\alpha=1.2$ and $\eta=4$ in (a), $\alpha=2$ and $\beta=0.5$ in (b), $\beta=0.1$ and $\eta=2.7$ in (c), $\gamma=2$ and $\eta=3$ in (d), $\gamma=2$ and $\alpha=1.01$ in (e) and $\gamma=2$ and $\beta=0.9$ in (f). The baby blue region represents the case that $\partial I$ is outside the horizon, the light yellow region represents the case that $\partial I$ is inside the horizon, and the dividing line between the baby blue region and the light yellow region represents the case that $\partial I$ is at the horizon.  }\label{region}
    \end{figure}

Combining the dimensionless parameters (\ref{59}) that we have defined, then we can reduce the inequalities (\ref{57}) and (\ref{58}) to simpler constraint conditions:
\begin{equation}\label{60}
2+\beta-\gamma>0, \; \frac{2e^{\gamma}(2+\beta-\gamma)}{(1-\alpha)(1+\beta-\gamma)}>0, \; \eta>\gamma>0, \;\alpha>0, \;\beta>0.
\end{equation}
Plug eq.(\ref{53}) for $v_I$ and the dimensionless parameters (\ref{59}) back to the term $8M-v_A+v_I$, then the inequalities (\ref{56}) can be reduced to
\begin{eqnarray}\label{61}
 2-\eta+\ln\left[\frac{2e^{\gamma}(2+\beta-\gamma)}{(1-\alpha)(1+\beta-\gamma)}\right]\begin{cases} <0, & u_I > u_H , \; \text{($\partial I$ is inside the horizon)}\\ =0, & u_I = u_H, \; \text{($\partial I$ is at the horizon)} \\ >0, &u_I > u_H, \; \text{($\partial I$ is outside the horizon)}\end{cases}
\end{eqnarray}

Combining the constraint conditions (\ref{60}) and the inequalities (\ref{61}), then we can plot out the relation between the changes of the dimensionless parameters and the distribution region of $\partial I$. We fix two dimensionless parameters in the inequalities (\ref{61}), then we can see the relation between the changes of the remaining two dimensionless parameters and three cases for the location of $\partial I$. The results are shown in Fig.\ref{region}.

In Fig.\ref{region}, there are six block diagrams such as $\alpha=1.2$ and $\eta=4$ in (a), $\alpha=2$ and $\beta=0.5$ in (b), $\beta=0.1$ and $\eta=2.7$ in (c), $\gamma=2$ and $\eta=3$ in (d), $\gamma=2$ and $\alpha=1.01$ in (e) and $\gamma=2$ and $\beta=0.9$ in (f). The baby blue region represents the case that $\partial I$ is outside the horizon, the light yellow region represents the case that $\partial I$ is inside the horizon, and the dividing line between the baby blue region and the light yellow region represents the case that $\partial I$ is at the horizon. The blank regions are excluded.

\subsection{The observer in region III with $u_A > u_H$ }\label{666}

\begin{figure}
\centering
\includegraphics[scale=0.550]{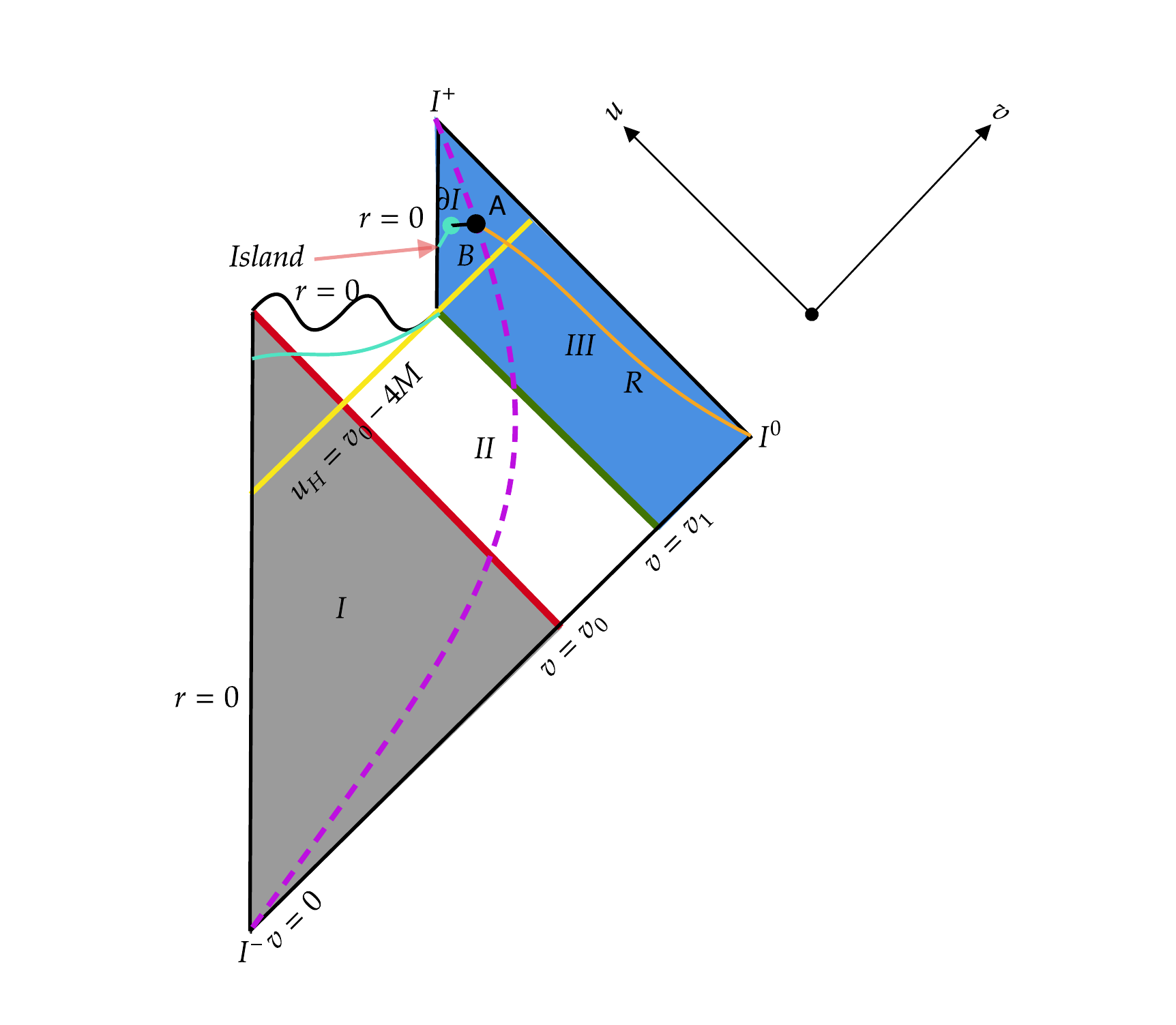}
\caption{The Penrose diagram for the step-function Vaidya model of an evaporating black hole spacetime with island. The observer is located in region III with $u_A > u_H$ after the black hole evaporates completely.} \label{o3}
\end{figure}

In this section, we consider the case that the observer on the cutoff surface $A$ is located in the Minkowski region III with $u_A > u_H$ after the black hole evaporates completely (see the Fig.\ref{o3}). In this case, $\partial I$ is only in region III with $u_I > u_H$.

In the Minkowski region III, the renormalized entanglement entropy of Hawking radiation is given by the eq.(\ref{26}). Plug eq.(\ref{26}) back to eq.(\ref{33}), then the generalized entropy of Hawking radiation in region III can be expressed as
\begin{equation}\label{62}
S_{\text{gen}}=\frac{\text{Area}(\partial I)}{4G_N}-\kappa c \frac{\text{ Area }}{L^2}=\frac{\pi r_{(I)}^2}{G_N}-\kappa c \frac{4\pi r_{(A)}^2}{L^2}.
\end{equation}
In order to calculate the generalized entropy of Hawking radiation for convenience, next we will choose $(U,v)$ coordinates in Minkowski region III. In flat spacetime, the geodesic distance $L$ between $\partial I$ and the cutoff surface $A$ is given by
\begin{equation}\label{63}
L(A,\partial I)=\sqrt{[U_I-U_A][v_A-v_I]}.
\end{equation}
Using the eq.(\ref{5}), then the expressions for $r_{(I)}$ and $r_{(A)}$ in $(U,v)$ coordinates are given by
\begin{equation}\label{64}
r_{(I)}=\frac{v_I-U_I}{2}, \; r_{(A)}=\frac{v_A-U_A}{2}.
\end{equation}
The region $B$ is a piece of Cauchy surface, which is a spacelike surface with $ds^2(A,\partial I)>0$. Then the square of the geodesic distance between $\partial I$ and the cutoff surface $A$ is greater than zero:
\begin{equation}\label{76}
L^2(A,\partial I)=[U_I-U_A][v_A-v_I]>0.
\end{equation}
From the eq.(\ref{76}), we have
\begin{eqnarray}\label{77}
\begin{cases} v_I < v_A, & U_I >U_A , \; \Longrightarrow v_I-U_I<v_A-U_A, \\ v_I > v_A, & U_I <U_A, \; \Longrightarrow v_I-U_I>v_A-U_A.\end{cases}
\end{eqnarray}
Combining eq.(\ref{64}) and the inequalities (\ref{77}), we can confirm that
\begin{eqnarray}
\begin{cases} v_I < v_A, & U_I >U_A , \; \Longrightarrow r_{(I)}<r_{(A)}, \\ v_I > v_A, & U_I <U_A, \; \Longrightarrow r_{(I)}>r_{(A)}.\end{cases}
\end{eqnarray}

Put the eqs.(\ref{63}) and (\ref{64}) into the eq.(\ref{62}), then the generalized entropy $S_{\text{gen}}$ of Hawking radiation can be written as
\begin{equation}\label{65}
S_{\text{gen}}=\frac{\pi(v_I-U_I)^2}{4G_N}- \frac{\kappa c\pi(v_A-U_A)^2}{(U_I-U_A)(v_A-v_I)}.
\end{equation}
Extremizing $S_{\text{gen}}$ (\ref{65}) over $U_I$ and $v_I$ respectively, then we have
\begin{eqnarray}
 \frac{\partial S_{\text{gen}}}{\partial U_I} &=& \frac{\pi(U_I-v_I)}{2G_N}+ \frac{\kappa c\pi(U_A-v_A)^2}{(U_A-U_I)^2(v_A-v_I)}=0 \label{66},\\
\frac{\partial S_{\text{gen}}}{\partial v_I}&=&\frac{\pi(v_I-U_I)}{2G_N}+ \frac{\kappa c\pi(U_A-v_A)^2}{(U_A-U_I)(v_A-v_I)^2}=0\label{67}.
 \end{eqnarray}
It is easy to see that we can simplify the eqs.(\ref{66}) and (\ref{67}) as a simpler equation by adding them:
\begin{equation}\label{68}
\frac{\partial S_{\text{gen}}}{\partial U_I}+\frac{\partial S_{\text{gen}}}{\partial v_I}=\frac{\kappa c\pi(U_A-v_A)^2(v_A-v_I+U_A-U_I)}{(U_A-U_I)^2(v_A-v_I)^2}=0.
\end{equation}
From the eq.(\ref{68}) we can obtain
\begin{equation}\label{69}
v_A-v_I+U_A-U_I=0, \; \Longrightarrow v_A+U_A=v_I+U_I.
\end{equation}
Finally we can get $t_{(I)}=t_{(A)}$ by using $U+v=2t_{III}$, which means that $\partial I$ is located on an equal time Cauchy surface with the observer $A$.

To obtain the radial coordinate $r_{(I)}$ of $\partial I$, first we can transform eq.(\ref{67}) into a simple equation with a common denominator:
\begin{equation}\label{70}
\frac{\pi(v_I-U_I)(U_A-U_I)(v_A-v_I)^2+2G_N\kappa c\pi(U_A-v_A)^2}{2G_N(U_A-U_I)(v_A-v_I)^2}=0.
\end{equation}
From the eq.(\ref{70}), we can easily obtain
\begin{equation}\label{71}
\pi(v_I-U_I)(U_A-U_I)(v_A-v_I)^2+2G_N\kappa c\pi(U_A-v_A)^2=0.
\end{equation}
Using the first result $t_{(I)}=t_{(A)}$ that we have obtained, then we have
\begin{equation}\label{72}
v_I-U_I=2r_{(I)}, \; U_A-U_I=r_{(I)}-r_{(A)}, \; v_A-v_I=r_{(A)}-r_{(I)}, \; U_A-v_A=-2r_{(A)}.
\end{equation}
Put the eq.(\ref{72}) into the eq.(\ref{71}), then we can get a quartic equation with the variable $r_{(I)}$:
\begin{equation}\label{73}
2r_{(I)}\cdot(r_{(I)}-r_{(A)})^3+8G_N\kappa c\cdot r_{(A)}^2=0.
\end{equation}
There are four roots in the quartic equation (\ref{73}) more generally. The radial coordinate $r_{(I)}$ is a real number, so we should abandon the imaginary roots and remain the real root in eq.(\ref{73}). Next we will find out what the real root with physical significance is.

We plug the eq.(\ref{72}) back to the eq.(\ref{65}), then the generalized entropy $S_{\text{gen}}$ can be given by
\begin{equation}\label{74}
S_{\text{gen}}=\frac{\pi r_{(I)}^2}{G_N}- \frac{4\kappa c\pi r_{(A)}^2}{(r_{(A)}-r_{(I)})^2}.
\end{equation}
Compare the quartic equation with $r_{(I)}$ (\ref{73}) with the eq.(\ref{74}), it is easy to see that we can translate the expression for the generalized entropy $S_{\text{gen}}$ (\ref{74}) into a simpler expression without the term $\kappa c$:
\begin{equation}\label{75}
S_{\text{gen}}=\frac{\pi r_{(I)}(2r_{(I)}-r_{(A)})}{G_N}.
\end{equation}
The generalized entropy $S_{\text{gen}}$ is actually greater than zero or equal to zero, then we have $r_{(I)}\geq\frac{r_{(A)}}{2}$ from the eq.(\ref{75}). Combining the result $r_{(I)}<r_{(A)}$ or $r_{(I)}>r_{(A)}$ above that we have confirmed, then we can further confirm that the range of $r_{(I)}$ is
\begin{equation}\label{78}
\frac{r_{(A)}}{2}\leq r_{(I)}<r_{(A)}.
\end{equation}

\begin{figure}
\centering
\includegraphics[scale=0.750]{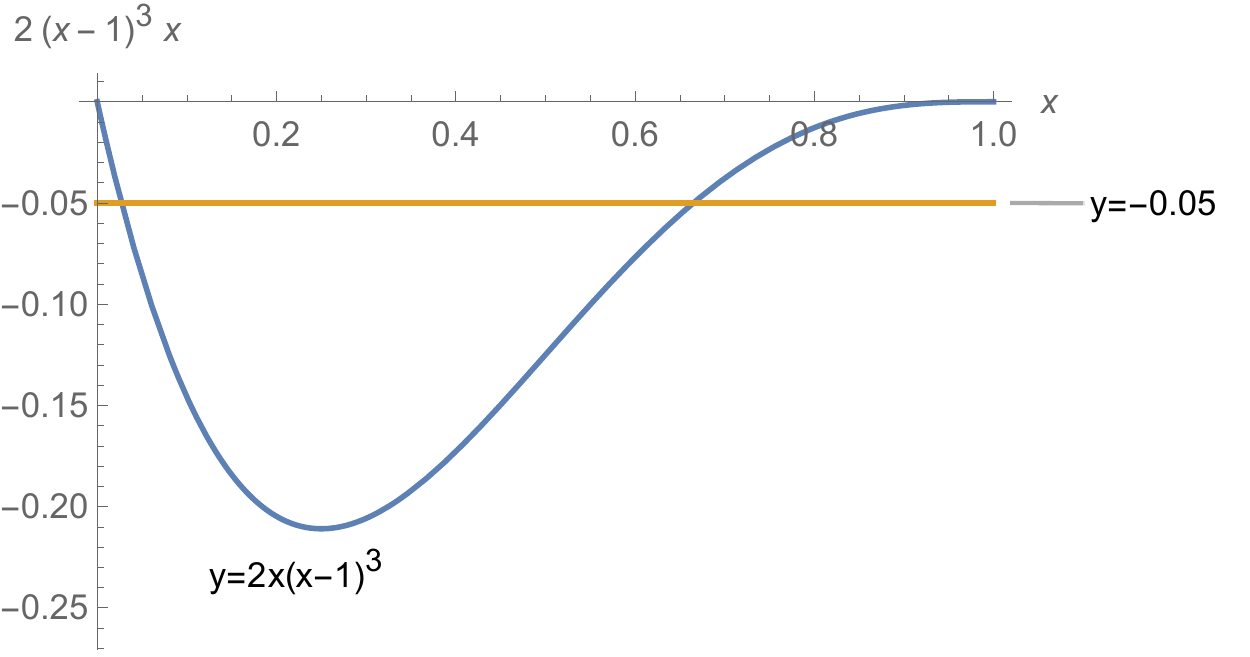}
\caption{The curve $y=2x\cdot(x-1)^3$ and the line $y=-0.05$ with the domain of definition $x\in (0,1)$ with $\frac{8G_N\kappa c }{r_{(A)}^2}=0.05$.  } \label{Re3}
\end{figure}

To reduce the quartic equation (\ref{73}) to a simpler equation, first we define a dimensionless parameter as $x\equiv\frac{r_{(I)}}{r_{(A)}}$ with the domain of definition $x\in[\frac{1}{2},1)$. Then the quartic equation (\ref{73}) becomes
\begin{equation}\label{79}
2x\cdot(x-1)^3+\frac{8G_N\kappa c }{r_{(A)}^2}=0, \; \Longrightarrow 2x\cdot(x-1)^3=-\frac{8G_N\kappa c }{r_{(A)}^2}.
\end{equation}
We define a dimensionless parameter as $\xi\equiv-\frac{8G_N\kappa c }{r_{(A)}^2}$, then we can see that the line $y=\xi$ intersects the curve $y=2x\cdot(x-1)^3$ at two points which correspond to two real roots in the eq.(\ref{73}), but there is only a point of intersection that is in the domain of definition $x\in[\frac{1}{2},1)$. The result is shown in Fig.\ref{Re3}. In the domain of definition $x\in[\frac{1}{2},1)$, we can determine that the range of $\xi$ is $-\frac{1}{8}\leq\xi<0$ from the eq.(\ref{79}).    At last, then we can confirm that there exists an island in the Minkowski region III after the black hole evaporates completely when $r_{(A)}^2\geq64G_N\kappa c $.

\section{Conclusion and discussion}\label{777}

In this paper, we apply the semiclassical method of quantum extremal surface (QES) to solve the information paradox for the step-function Vaidya model of evaporating black holes. Based on the motivation, we choose the ``in'' vacuum state to describe the black hole produced by the dynamical gravitational collapse, since it contains no incoming thermal radiation coming in from the past null infinity $\mathcal{J}^-$. Then the entanglement entropy of Hawking radiation in $4D$ dynamical spherically symmetric spacetime can be calculated by $s$-wave approximation for a distant observer.

Up to now, it is well known that most studies on the information problem have been concentrated to eternal black holes with $D\geq3$ \cite{Saha:2021ohr,Yu:2021rfg,Hashimoto:2020cas,Matsuo:2020ypv,Arefeva:2021kfx,Wang:2021woy,Kim:2021gzd}. In our previous work \cite{Gan:2022jay}, unlike other work, we have studied and discussed the Page curve and island for one-sided asymptotically flat black hole. In this paper, we take one more step further by analyzing and studying the entanglement islands for the step-function Vaidya model of evaporating black holes at very late times, which has a Minkowski region III more than the one-sided asymptotically flat black hole.

For the whole spacetime of the step-function Vaidya model of evaporating black holes, we have got the differential relations of the coordinate that matches across the ingoing null geodesics $v=v_0$ and $v=v_1$. Then we solve the differential relations and get the spacetime metrics of the three regions in $(u,v)$ coordinates, which is convenient to calculate the generalized entropy of Hawking radiation. We find that, when the observer on the cutoff surface $A$ is located in the Minkowski region III with $u_A < u_H$ at very late times, the location of $\partial I$ depends on the value of $8M-v_A+v_I$. $\partial I$ is inside the horizon with $8M-v_A+v_I<0$, at the horizon with $8M-v_A+v_I=0$ and outside the horizon with $8M-v_A+v_I>0$. Moreover, when the observer on the cutoff surface $A$ is located in the Minkowski region III with $u_A > u_H$ after the black hole evaporates completely, we find that there exists an island in the Minkowski region III and $\partial I$ is located on an equal time Cauchy surface with the observer $A$ when $r_{(A)}^2\geq64G_N\kappa c $.

\section*{Acknowledgments}

This work is supported by the National Natural Science Foundation of China under Grant No. 11975116.

\begin{appendix}
\section{The stress-energy tensor of the matter field in the step-function Vaidya model of an evaporating black hole }\label{88838}

The step-function Vaidya model of an evaporating black hole can partially model the evaporation of a black hole, and the stress-energy tensor $\langle T_{\mu\nu}\rangle$ of a quantized massless scalar field can be calculated explicitly for the entire spacetime. The spacetime's metric for the three regions can be written in terms of the double-null coordinates as the eq.(\ref{88386}). From the above eqs.(\ref{3}) and (\ref{13}) that we have obtained, we can summarize the spacetime's metric of the three regions in terms of the $(u,v)$ coordinates:
\begin{eqnarray}
 ds^2 &=& -dudv, \quad v \leq v_0,\\
 ds^2 &=&-\left(1-\frac{2M}{r_{II}}\right)\frac{u-v_0}{u+4M-v_0}dudv, \quad v_0<v<v_1,\\
 ds^2 &=&-\frac{u-v_0}{u+4M-v_0}\cdot\frac{v_1-U-4M}{v_1-U}dudv, \quad v_1<v.
\end{eqnarray}

The stress-energy tensor $\langle T_{\mu\nu}\rangle$ of a quantized massless scalar field for the entire spacetime can be calculated by quantum field theory in curved spacetime. In the Minkowski region I, the stress-energy tensor is
\begin{equation}
T_{\mu\nu}=0, \quad v \leq v_0.
\end{equation}
In the Schwarzschild region II $(v_0<v<v_1)$, the stress-energy tensor was firstly calculated by Davies, Fulling and Unruh \cite{Davies:1976ei}:
\begin{eqnarray}
 T_{u^\ast u^\ast} &=& (24\pi)^{-1}(-\frac{M}{r_{II}^3}+\frac{3}{2}\frac{M^2}{r_{II}^4}-\frac{8M}{u^3}-\frac{24M^2}{u^4}),\\
 T_{vv} &=& (24\pi)^{-1}(-\frac{M}{r_{II}^3}+\frac{3}{2}\frac{M^2}{r_{II}^4}),\\
 T_{u^\ast v} &=& (24\pi)^{-1}(-\frac{M}{r_{II}^3})(1-\frac{2M}{r_{II}}).
\end{eqnarray}
The region II in this toy model is exactly Schwarzschild region until the moment of evaporation $v=v_1$, and we can see that $\langle T_{\mu\nu}\rangle$ is finite and regular everywhere except at the curvature singularity in region II. In the Minkowski region III $(v_1<v)$, the stress-energy tensor of the quantized massless scalar field in this model was firstly calculated by Hiscock \cite{Hiscock:1980ze}:
\begin{eqnarray}
 T_{vv} &=& T_{Uv}=0,\\
 T_{UU} &=& \frac{M}{3\pi(v_1-U-4M)^2}(\frac{3M}{(v_1-U)^2}+\frac{1}{(v_1-U)}-\frac{(v_1-U)^2}{u^3}-\frac{3M(v_1-U)^2}{u^4})\label{38688}.
\end{eqnarray}
The null hypersurface $U=v_1-4M$ is the continuation of the apparent horizon $r_{II}=2M$ into the region III, and the limiting value of $T_{UU}$ in the eq.(\ref{38688}) on the apparent horizon (extended into the region III) is
\begin{equation}\label{477}
\lim_{U\rightarrow v_1-4M}T_{UU}=(128\pi M^2)^{-1}(1-e^{-\frac{v_1}{2M}}).
\end{equation}
From the eq.(\ref{38688}), it is easy to see that the value of $T_{UU}$ is divergent at the Cauchy horizon $(U=v_1)$. The energy density diverges as $(v_1-U)^{-2}$, and then the integrated energy outflow diverges as $(v_1-U)^{-1}$. This is a positive energy divergence, and we can expand $T_{UU}$ to the
first order of $(v_1-U)^{-1}$ near $U=v_1$:
\begin{equation}\label{477}
T_{UU}=(16\pi)^{-1}(v_1-U)^{-2}+O((v_1-U)^{-1}).
\end{equation}

Since $T_{u^\ast u^\ast}$ is regular along the event horizon, then we can conclude that the infinite flux of outgoing Hawking radiation is produced by the naked singularity.

\end{appendix}

\end{document}